\documentclass[a4paper,fleqn]{cas-sc}

\usepackage[numbers]{natbib}
\usepackage{enumitem}
\usepackage{amsmath,amssymb,amsfonts}
\usepackage{algorithm}
\usepackage{algpseudocode}
\usepackage{subcaption}

\begin{document}
\let\WriteBookmarks\relax
\def\floatpagepagefraction{1}
\def\textpagefraction{.001}

\shorttitle{Meta Learners for Fundus Image Segmentation}

\shortauthors{P.A.~Zumarsyah et~al.}

\title [mode = title]{Meta-learners for few-shot weakly-supervised optic disc and cup segmentation on fundus images}

\author[1]{Pandega Abyan Zumarsyah}[orcid=0009-0004-0018-6252]
\ead{pandegaabyanzumarsyah@mail.ugm.ac.id}

\author[1]{Igi Ardiyanto}[orcid=0000-0002-0006-1458]
\ead{igi@ugm.ac.id}

\author[1]{Hanung Adi Nugroho}[orcid=0000-0001-7749-8044]
\cormark[1]
\ead{adinugroho@ugm.ac.id}

\affiliation[1]{organization={Department of Electrical and Information Engineering, Faculty of Engineering, Universitas Gadjah Mada},
	addressline={Grafika~St. No.~2},
	city={Yogyakarta},
	postcode={55281},
	country={Indonesia}}

\cortext[cor1]{Corresponding author}

\begin{abstract}
	This study develops meta-learners for few-shot weakly-supervised segmentation (FWS) to address the challenge of optic disc (OD) and optic cup (OC) segmentation for glaucoma diagnosis with limited labeled fundus images. 
	We significantly improve existing meta-learners by introducing Omni meta-training which balances data usage and diversifies the number of shots. 
	We also develop their efficient versions that reduce computational costs. 
	In addition, we develop sparsification techniques that generate more customizable and representative scribbles and other sparse labels.
	After evaluating multiple datasets, we find that Omni and efficient versions outperform the original versions, with the best meta-learner being Efficient Omni ProtoSeg (EO-ProtoSeg).
	It achieves intersection over union (IoU) scores of 88.15\% for OD and 71.17\% for OC on the REFUGE dataset using just one sparsely labeled image, outperforming few-shot and semi-supervised methods which require more labeled images. 
	Its best performance reaches 86.80\% for OD and 71.78\% for OC on DRISHTI GS, 88.21\% for OD and 73.70\% for OC on REFUGE, 80.39\% for OD and 52.65\% for OC on REFUGE. 
	EO-ProtoSeg is comparable to unsupervised domain adaptation methods yet much lighter with less than two million parameters and does not require any retraining.
\end{abstract}


\maketitle

\section{Introduction}

Glaucoma, the second leading cause of blindness, affected 80 million people in 2020 \cite{zedanAutomatedGlaucomaScreening2023} and is projected to affect 112 million by 2040 \cite{raveendranCurrentInnovationsIntraocular2023}.
Its symptoms may go unnoticed by patients for months or even years after damage occurs \cite{maReviewOpticDisc2024}.
Moreover, currently there is no treatment to reverse the damage caused by glaucoma \cite{kashyapGlaucomaDetectionClassification2022}.
However, early diagnosis can prevent the damage from worsening \cite{xiongWeakLabelBased2022,zedanAutomatedGlaucomaScreening2023}.
The diagnosis can be done by analyzing the retinal area through noninvasive fundus imaging \cite{zedanAutomatedGlaucomaScreening2023}.
The analysis includes segmentation of some parts of fundus images, namely the optic disc (OD) and optic cup (OC) \cite{kashyapGlaucomaDetectionClassification2022,xiongWeakLabelBased2022,gargRealTimeCloudbased2022,zhaoApplicationAttentionUNet2021}.
However, the analysis involving segmentation by experts is often complicated, costly, and time-consuming \cite{zedanAutomatedGlaucomaScreening2023, xiongWeakLabelBased2022, gargRealTimeCloudbased2022,maReviewOpticDisc2024}.

Addressing the issue, deep learning has been used for automatic OD and OC segmentation \cite{zedanAutomatedGlaucomaScreening2023,maReviewOpticDisc2024,alawadMachineLearningDeep2022}.
Many previous works have achieved high accuracy by utilizing advanced techniques like transformer \cite{yiC2FTFNetCoarsetofineTransformer2023}, patch-based network \cite{pandaGlaucoNetPatchBasedResidual2021}, or graph convolutional network \cite{tianGraphConvolutionalNetwork2020}.
Despite the high accuracy, deep learning requires many images that are fully annotated down to each of their pixels \cite{zedanAutomatedGlaucomaScreening2023,xiongWeakLabelBased2022,maReviewOpticDisc2024,alawadMachineLearningDeep2022}.
They are called dense labels and obtaining them for numerous fundus images is costly and time-consuming \cite{xiongWeakLabelBased2022,maReviewOpticDisc2024}.
The limited availability of labeled data is challenging for common deep learning approaches \cite{tajbakhshEmbracingImperfectDatasets2020,zhaoLEUDALabelEfficientUnsupervised2023,zhangCollaborativeUnsupervisedDomain2020}.
Therefore, common deep learning approaches are less suitable for OD and OC segmentation due to the limited availability of densely labeled fundus images.

To solve this problem, there are various approaches to adapt deep learning for OD and OC segmentation with limited labeled data.
One of the most common and straightforward approaches is transfer learning (TL) \cite{liuReviewSelfsupervisedGenerative2024}.
TL involves training the model on large and easily accessible datasets (source), then fine-tuning it on smaller and problem-specific datasets (target) \cite{zhaoApplicationAttentionUNet2021}.
However, recent works using TL still require many labeled target images \cite{zhaoApplicationAttentionUNet2021, kashyapGlaucomaDetectionClassification2022}.
Additionally, TL is known to be less effective when the target dataset is either too different or too limited \cite{liFewShotDomainAdaptation2021, wangMedicalImageSegmentation2022}.

Another common approach for OD and OC segmentation is unsupervised domain adaptation (UDA).
It adapts a model from a source domain with fully labeled images to a target domain without labeled images.
UDA usually involve adversarial learning \cite{liuCADAMultiscaleCollaborative2022,zhangConvolutionalAutoencoderJoint2022,zhouUnsupervisedDomainAdaptation2024,chenIOSUDAUnsupervisedDomain2021,heSelfensemblingMaskboundaryDomain2024,wangPatchBasedOutputSpace2019,bianUnsupervisedOpticDisc2019,liuECSDNetJointOptic2022,xuMinimizingEntropyFourierConsistency2021,kadambiWGANDomainAdaptation2020,zhangUnsupervisedDomainAdaptation2022} with segmentor and discriminator networks.
Some works also involve consistency loss \cite{heSelfensemblingMaskboundaryDomain2024,xuMinimizingEntropyFourierConsistency2021}, mean teacher networks \cite{liuCADAMultiscaleCollaborative2022,heSelfensemblingMaskboundaryDomain2024}, and image synthesis \cite{zhangConvolutionalAutoencoderJoint2022,chenIOSUDAUnsupervisedDomain2021,bianUnsupervisedOpticDisc2019,zhangUnsupervisedDomainAdaptation2022}.
However, UDA methods generally need access to both source and target datasets for training \cite{zhangRobustColorMedical2022,huDevilChannelsContrastive2023}.
The network also needs full retraining each time the target domain is changed \cite{wangPatchBasedOutputSpace2019}.

An alternative to UDA is few-shot segmentation (FSS) that utilizes a few densely labeled target images without full retraining.
However, it is less common in OD and OC segmentation.
Li et al. \cite{liFewShotDomainAdaptation2021} propose a Polymorphic Transformer that uses source datasets to update its parameters and then uses target datasets to update its projection parameters.
Some studies also perform semi-supervised segmentation (SSS), which involves a small number of labeled images and a large number of unlabeled images.
Wu et al. \cite{wuMinimizingEstimatedRisks2022} develop an SSS method that involves a modified UNet and a risk estimator on unlabeled images.
Tang et al. \cite{tangConsistencyAdversarialSemisupervised2023} also perform SSS, but use mean teacher networks with adversarial learning.
Meanwhile, Mei et al. \cite{meiSemisupervisedImageSegmentation2024} introduce residual perturbation and exponential Dice loss in addition to the mean teacher networks.
Due to the usage of a few labeled target images, studies by Wu et al. \cite{wuMinimizingEstimatedRisks2022} and Tang et al. \cite{tangConsistencyAdversarialSemisupervised2023} can also be considered as FSS.

Instead of using a small number of dense labels, some works use weakly-supervised segmentation (WSS) that utilizes a large number of weak labels.
Most weak labels used in OD and OC segmentation are image-level classes \cite{luWeaklySupervisedSemantic2019} and bounding boxes \cite{luWeaklySupervisedSemiSupervised2020,luWeaklySupervisedSemantic2019,wangCDRNetAccurateCuptodisc2023}.
Most of these works generate pseudo-labels for supervised learning that involve GrabCut \cite{luWeaklySupervisedSemantic2019,luWeaklySupervisedSemiSupervised2020} or multi-instance learning \cite{wangCDRNetAccurateCuptodisc2023}.
Another type of weak labels is sparse labels where only a few pixels in the image are labeled with points \cite{ouassitBriefSurveyWeakly2022, zhuSurveyWeaklysupervisedSemantic2023}, scribbles \cite{shenSurveyLabelEfficientDeep2023, ouassitBriefSurveyWeakly2022, zhuSurveyWeaklysupervisedSemantic2023}, or other forms.
However, sparse labels are less common in OD and OC segmentation.

The combination of FSS and WSS is the few-shot weakly-supervised segmentation (FWS).
It utilizes a few sparsely labeled images using a method called meta-learning.
Compared to UDA, FWS does not require full retraining each time the target domain changes.
Unlike FSS/SSS and WSS, FWS does not need dense labels or a large number of labeled images.
Although promising, we found little to no published works that perform FWS for OD and OC on fundus images, except our preliminary work \cite{zumarsyahFewShotWeaklySupervised2024}.

There are some novel meta-learners which are used for FWS: Holistic Prototype Activation (HPA) \cite{chengHolisticPrototypeActivation2023}, Base and Meta (BAM) \cite{wangPANetFewShotImage2019}, Weakly-supervised Segmentation Learning (WeaSeL) \cite{gamaLearningSegmentMedical2021}, ProtoSeg \cite{gamaWeaklySupervisedFewShot2023}, and Pseudo-labeling Texture Segmentation (PTS) \cite{hanPseudolabelingBasedWeakly2024}.
HPA and BAM perform well on common images, but their architecture requires a model pre-trained on large datasets.
Thus, they are unsuitable for medical images due to the lack of pre-trained models and large datasets.
WeaSeL \cite{gamaLearningSegmentMedical2021} and ProtoSeg \cite{gamaWeaklySupervisedFewShot2023} have been comprehensively evaluated on various medical images, but are still limited to grayscale images and binary segmentation.
PTS \cite{hanPseudolabelingBasedWeakly2024} has been evaluated on fundus images, but the task is blood vessel binary segmentation, which is different from OD and OC segmentation.
Moreover, the results of PTS on fundus images are not satisfactory since PTS is designed mainly for material microstructure images.

In this work, the objective is to adapt and improve existing meta-learners, namely WeaSeL and ProtoSeg, for few-shot weakly-supervised OD and OC segmentation on fundus images.
WeaSeL and ProtoSeg were chosen because they have been evaluated on various medical images and have been shown to work well.
Meanwhile, HPA, BAM, and PTS are less suitable for medical images and harder to train due to their complex architectures.
In addition, we improved existing sparsification techniques, which simulate sparse labels from dense labels.
The contributions of this work are:
\begin{enumerate}[nosep]
	\item Improvement of sparsification techniques making them more realistic, customizable, and suitable for multiclass labels.
	\item Adaptation of WeaSeL and ProtoSeg to support multichannel images and multiclass labels.
	\item Performance improvement of WeaSeL and ProtoSeg by optimizing data utilization and introducing diverse numbers of shots during training.
	\item Inference time reduction of WeaSeL and ProtoSeg without performance loss.
	\item Comprehensive and consistent evaluations across multiple meta-learners, datasets, sparse label types, and numbers of shots.
\end{enumerate}

\section{Methodology}

\begin{figure}[htbp!]
	\centering
	\includegraphics[scale=0.3]{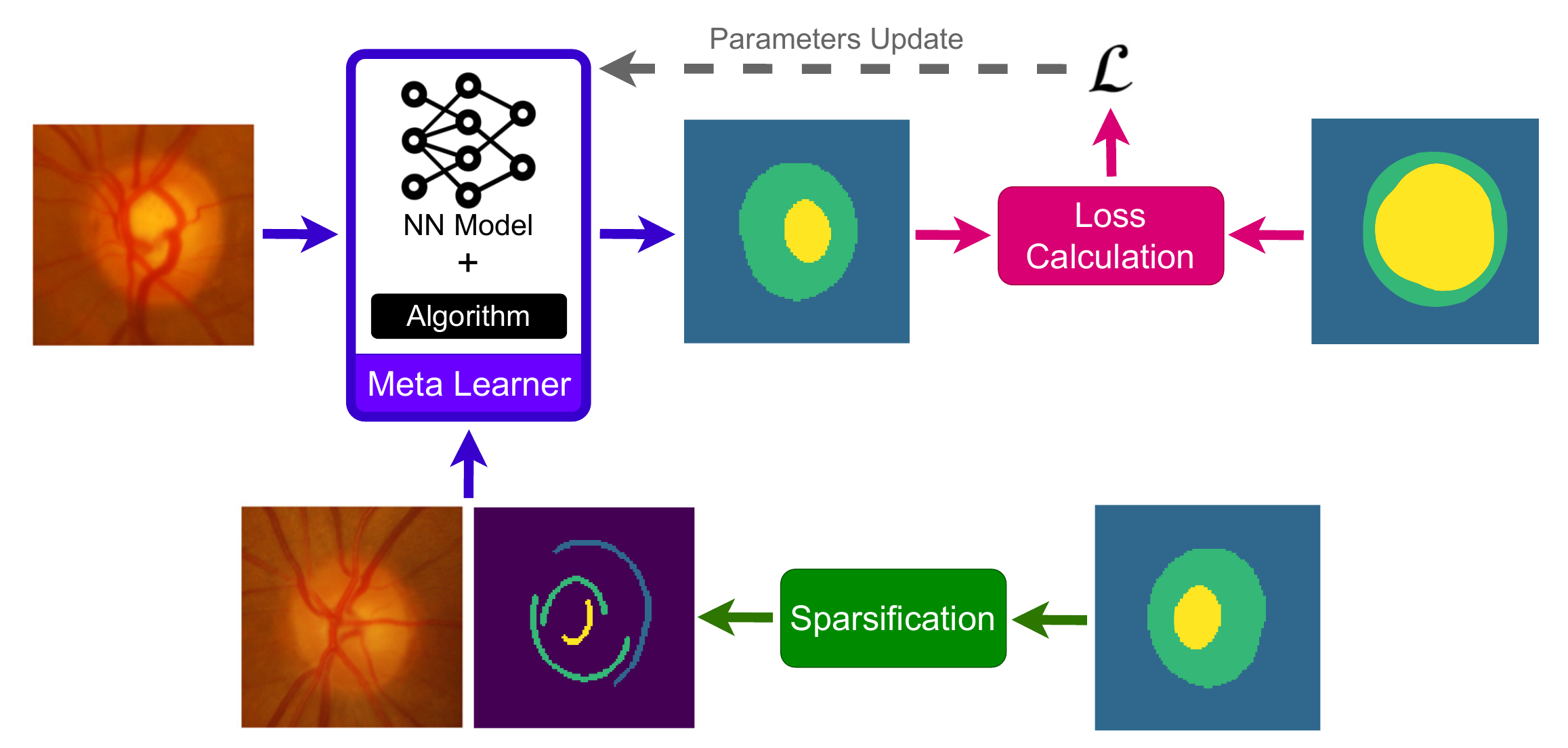}
	\caption{Simple system diagram of our work with sparsification and meta-learner as the main components.}
	\label{fig:system_diagram}
\end{figure}

Fig. \ref{fig:system_diagram} shows the simple system diagram of our work.
It consists of two main processes, the sparsification and the meta-learner.
There is also a query loss calculation, but it can be considered as part of the meta-learner and usually just similar to a cross-entropy loss.
Note that the sparsification and loss calculation are not used during inference.

Regarding the meta-learner, it takes query images, support images, and support sparse masks as input and returns segmentation predictions for the query images.
The meta-learner consists of a neural network model capable of performing basic image segmentation and an algorithm for utilizing the model.
In this case, WeaSeL and ProtoSeg are the algorithms since they are model-agnostic.

\subsection{Sparsification Techniques}

\begin{figure}[htbp!]
	\centering
	\includegraphics[scale=0.14]{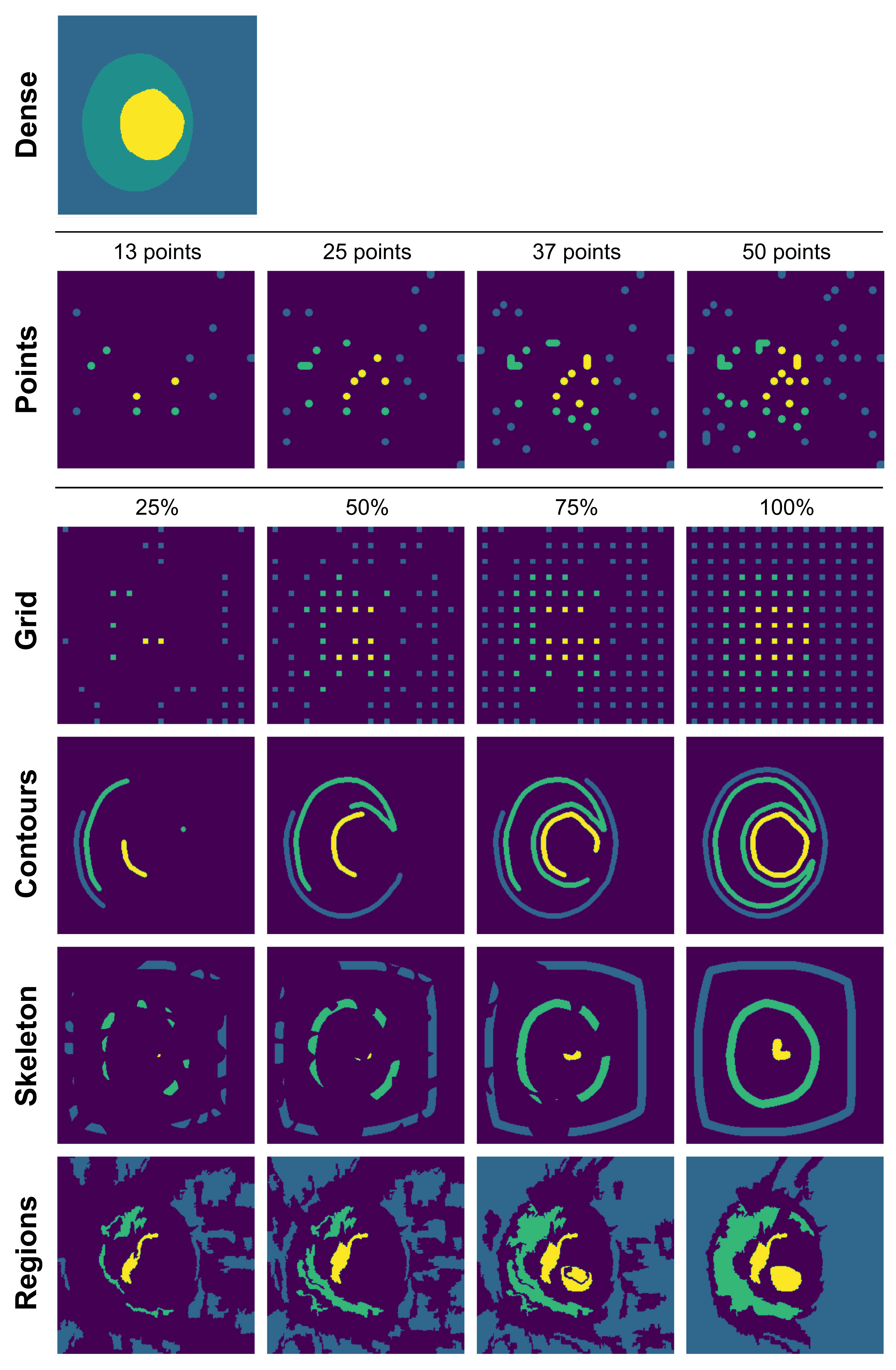}
	\caption{Various sparse labels generated from a dense label.
		The left axis indicates the sparsification technique while the top axis indicates the density value.
		The labels have four classes: OC (yellow), OD (green), background (teal blue), and unannotated (purple).
		Note that these are actual labels used in the experiments.}
	\label{fig:sparse_label}
\end{figure}

We developed five sparsification techniques: points, grid, contours, skeleton, and regions.
They are similar to those developed by Gama et al. \cite{gamaWeaklySupervisedFewShot2023}, but there are some modifications.
All of them are adapted for multiclass labels.
In addition, points and grid techniques are developed with customizable point sizes.
With a larger but appropriate point size, annotators can mark more pixels for each point using the same effort.
There are other parameters that can be customized for each technique.
Fig. \ref{fig:sparse_label} shows the actual outputs of each sparsification technique with various density values along with the dense label related to OD and OC segmentation.
Each sparsification technique takes some input, one of which is the density of the sparse label.
Then, some image processing steps are utilized to convert dense labels into sparse labels.

\subsubsection{Points}

It simulates an annotator marking some points in the image.
The inputs of this technique are a dense label $y$, the number of points $n_p$, and the size of each point $s_p$.
The $n_p$ is also the density parameter of this technique.
Meanwhile, the processing steps are:
\begin{enumerate}[nosep]
	\item Downscale $y$ to $s_p$ times smaller
	\item Select random $n_p$ pixels from downscaled $y$ as a sparse label
	\item Upscale the sparse label to the original size of $y$
	\item Apply morphological dilation to enhance the points in the sparse label
\end{enumerate}

\subsubsection{Grid}

It simulates an annotator marking some points provided in a grid pattern.
The inputs are $y$, the density or percentage of grid $p_g$, $s_p$, and the distance between points on the grid $s_g$.

Meanwhile, the processing steps of the grid technique are:

\begin{enumerate}[nosep]
	\item Downscale $y$ to $s_p$ times smaller
	\item Create a grid with $s_g$ pixels between points
	\item Select pixels in $y$ associated with points in the grid as a sparse label
	\item Upscale the sparse label to the original size of $y$
	\item Generate rounded blobs\footnote{https://scikit-image.org/docs/stable/api/skimage.data.html\#skimage.data.\\binary\_blobs} covering $p_s$ percent of the sparse label
	\item Filter out pixels of the sparse label that do not intersect with the blobs
\end{enumerate}

\subsubsection{Contours}

Contours represent the inner and outer boundaries of objects as drawn by an annotator.
Its inputs are $y$, the density or percentage of contours $p_c$, the radius of the disk to erode objects $s_e$, and the radius of the disk to dilate the sparse label $s_d$.
The processing steps are:

\begin{enumerate}[nosep]
	\item Erode objects in $y$ with a disk element of radius $s_e$
	\item Find contours on eroded objects using a marching squares algorithm\footnote{https://scikit-image.org/docs/stable/api/skimage.measure.html\#skimage.\\measure.find\_contours}
	\item Take $p_c$ percent of contours randomly and draw them as the sparse label
	\item Dilate the sparse label with a disk element of radius $s_d$
\end{enumerate}

\subsubsection{Skeleton}

Skeleton represents scribbles in the center of objects as drawn by an annotator.
Given $y$, density or percentage of skeleton $p_s$, and $s_d$, the processing steps are:

\begin{enumerate}[nosep]
	\item Generate the sparse label by applying a skeletonization algorithm\footnote{https://scikit-image.org/docs/stable/api/skimage.morphology.html\#skimage.\\morphology.skeletonize} on $y$
	\item Dilate the sparse label with a disk element of radius $s_d$
	\item Generate blobs and use them for filtering out the sparse label, like in the grid technique
\end{enumerate}

\subsubsection{Regions}

It simulates an annotator marking regions generated using SLIC (Simple Linear Iterative Clustering)\cite{achantaSLICSuperpixelsCompared2012}.
The inputs of this technique are $y$, the density or percentage of regions $p_r$, and the compactness of SLIC $s_c$.
The processing steps are:

\begin{enumerate}[nosep]
	\item Generate regions using SLIC with compactness $s_c$
	\item Filter out regions that contain multiple classes
	\item Select $p_r$ percent of filtered regions as the sparse label
\end{enumerate}

\subsection{Meta-Learner Algorithms}

\subsubsection{FWS Formulation}

Before discussing the algorithms of the meta-learners, clarifying the formulation of the FWS is essential.
This formulation follows \cite{gamaLearningSegmentMedical2021,gamaWeaklySupervisedFewShot2023} with some modifications.

A source dataset $\mathcal{S}$ consists of pairs $(x, y)$ where $x \in \mathbb{R}^{H \times W \times L}$ represents an image with dimensions $H \times W = N$ and $L$ channels.
Meanwhile, $y \in \mathbb{R}^{H \times W}$ is a dense segmentation label where each pixel in it has a value $c \in \{1, 2, \dots, C\}$, indicating the class label.
This dataset $\mathcal{S}$ is split into support $\mathcal{S}^{sup}$ and query $\mathcal{S}^{qry}$ such that $\mathcal{S}^{sup} \cap \mathcal{S}^{qry} = \emptyset$.

A target dataset $\mathcal{T}$ can also be split into support $\mathcal{T}^{sup}$ and query $\mathcal{T}^{qry}$.
The support $\mathcal{T}^{sup}$ is composed of pairs $(x^{sup}, \acute{y}^{sup})$ where $x \in \mathbb{R}^{H \times W \times L}$ is an image while $\acute{y} \in \mathbb{R}^{H \times W}$ is a sparse segmentation label.
Note that the support size $T = |\mathcal{T}^{sup}|$ is limited, typically fewer than 20.
Regarding the query $\mathcal{T}^{qry}$, it contains only unlabeled images $x^{qry}$.

During training, a meta-learner utilizes the source dataset $\mathcal{S}$ for optimizing a neural network $\Phi$ with parameters $\theta$.
During inference, it takes $x^{qry}$ and $\mathcal{T}^{sup}$ of the target dataset as input and outputs the predicted segmentation label $\hat{y}^{qry}$.
In this work, we develop and implement two meta-learners along with their improved versions: WeaSeL, ProtoSeg, Omni WeaSeL (O-WeaSeL), Omni ProtoSeg (O-ProtoSeg), Efficient Omni WeaSeL (EO-WeaSeL), and Efficient Omni ProtoSeg (EO-ProtoSeg).

\subsubsection{WeaSeL and ProtoSeg}

Algorithm \ref{alg:ori_meta_training} shows the original meta-training of WeaSeL and ProtoSeg as implemented by Gama et al. \cite{gamaWeaklySupervisedFewShot2023}.
Each iteration $i$ randomly selects $B$ images and their corresponding sparse labels from support $\mathcal{S}^{sup}$ with $B$ as the batch size.
The support images and sparse labels are then combined into batches $X^{sup}$ and $\acute{Y}^{sup}$.
It also randomly selects $B$ images and their corresponding dense labels from query $\mathcal{S}^{qry}$ to form batches $X^{qry}$ and $Y^{qry}$.
Note that the random selections are done with replacement.
The meta-training step then takes the four batches as input and outputs the loss $\mathcal{L}$.
The parameters $\theta$ of the neural network $\Phi$ are updated using an optimizer algorithm based on the gradient of the loss $\nabla_{\theta} \mathcal{L}$.
The detail of the meta-training step is shown in Algorithm \ref{alg:weasel_step} for WeaSeL and Algorithm \ref{alg:protoseg_step} for ProtoSeg.

\begin{algorithm}
	\caption{Original Meta-Training}
	\label{alg:ori_meta_training}
	\begin{algorithmic}
		\Require $\mathcal{S}$, $n_{ep}$, $n_{iter}$
		\State Initialize $\theta$
		\For{$e = 1$ to $n_{ep}$}
		\For{$i = 1$ to $n_{iter}$}
		\State Take $X^{sup}$, $\acute{Y}^{sup}$ randomly from $\mathcal{S}^{sup}$
		\State Take $X^{qry}$, $Y^{qry}$ randomly from $\mathcal{S}^{qry}$
		\State Run Step on $X^{sup}$, $\acute{Y}^{sup}$, $X^{qry}$, $Y^{qry}$ to obtain $\mathcal{L}$
		\State Update $\theta \leftarrow \theta - \text{opt}(\nabla_{\theta} \mathcal{L})$
		\EndFor
		\EndFor
	\end{algorithmic}
\end{algorithm}

\begin{algorithm}
	\caption{WeaSeL Meta-Training Step}
	\label{alg:weasel_step}
	\begin{algorithmic}
		\Require $X^{sup}$, $\acute{Y}^{sup}$, $X^{qry}$, $Y^{qry}$, $\acute{\eta}$
		\State Predict $\hat{Y}^{sup} = \Phi_{\theta}(X^{sup})$
		\State Compute $\mathcal{L}_{SCE}$ from $(\acute{Y}^{sup}, \hat{Y}^{sup})$ using \eqref{eq:sce_loss}
		\State Update $\theta \leftarrow \theta - \acute{\eta} \nabla_{\theta} \mathcal{L}_{SCE}$
		\State Predict $\hat{Y}^{qry} = \Phi_{\theta}(X^{qry})$
		\State Compute $\mathcal{L}_{CE}$ from $(Y^{qry}, \hat{Y}^{qry})$ using \eqref{eq:ce_loss}
		\State \Return $\mathcal{L}_{CE}$
	\end{algorithmic}
\end{algorithm}

\begin{algorithm}
	\caption{ProtoSeg Meta-Training Step}
	\label{alg:protoseg_step}
	\begin{algorithmic}
		\Require $X^{sup}$, $\acute{Y}^{sup}$, $X^{qry}$, $Y^{qry}$
		\State Obtain $E^{sup}$ from $X^{sup}$ using \eqref{eq:embedding}
		\State Compute $\mathcal{P}_c$ for each $c$ from $(E^{sup}, \acute{Y}^{sup})$ using \eqref{eq:cls_prototype}
		\State Compute $\rho_c$ for each $c$ from $(X^{qry}, \mathcal{P}_c)$ using \eqref{eq:cls_probability}
		\State Compute $\mathcal{L}_{PRO}$ from $(Y^{qry}, \rho_c)$ using \eqref{eq:proto_loss}
		\State \Return $\mathcal{L}_{PRO}$
	\end{algorithmic}
\end{algorithm}

The meta-training step of WeaSeL starts by predicting the support labels $\hat{Y}^{sup}$ using the neural network $\Phi$.
It then computes a cross-entropy loss designed to handle sparse labels:

\begin{equation}
	\label{eq:sce_loss}
	\mathcal{L}_{SCE} = - \frac{1}{N} \sum_{j}^{N} \sum_{i}^{C} w^j \acute{Y}_i^j \log(\hat{Y}_i^j)
\end{equation}

In Equation \eqref{eq:sce_loss} and the following equations, $N = H \times W$ is the number of pixels in the image, $j$ in the superscript is the pixel index, while $i$ in the subscript indicates the class index.
Meanwhile, $w^j$ is the weight of the pixel $j$, which is $1$ if the pixel is annotated and $0$ otherwise.

The computed loss is then used to update the neural network $\Phi$ based on the gradient of the loss and an inner learning rate $\acute{\eta}$.
After that, the updated neural network is used to predict the query labels $\hat{Y}^{qry}$.
The cross-entropy loss is then computed for the query labels:

\begin{equation}
	\label{eq:ce_loss}
	\mathcal{L}_{CE} = - \frac{1}{N} \sum_{j}^{N} \sum_{i}^{C} Y_i^j \log(\hat{Y}_i^j)
\end{equation}

The loss is then returned to the meta-training algorithm for backpropagation.
Note that this backpropagation is very deep since the forward pass involves two neural network predictions, making the training process require more time and memory.
During inference, the neural network $\Phi$ is tuned on the support for some epochs using standard learning, and then it can be used to predict the query just like a standard neural network.

Regarding ProtoSeg, the meta-training step starts by obtaining the support embeddings $E^{sup}$ of size or depth $M$ using a neural network $\Phi$:

\begin{equation}
	\label{eq:embedding}
	E = \Phi_{\theta}^{M} (X)
\end{equation}

It then used to compute the class prototypes $\mathcal{P}_c$ for each class $c$ using the support sparse labels $\acute{Y}^{sup}$:

\begin{equation}
	\label{eq:cls_prototype}
	\mathcal{P}_c = \frac{1}{N_c} \sum_{j}^N Y_c^j \odot E^{sup,j}
\end{equation}

In Equation \eqref{eq:cls_prototype}, $N_c$ is the number of annotated pixels for class $c$ while $\odot$ is the element-wise multiplication.
After that, the query embeddings are obtained using Equation \eqref{eq:embedding} and compared with the support prototypes using an Euclidean distance function $d()$.
Softmax is then applied to the distances to obtain the class probabilities $\rho_c$:

\begin{equation}
	\label{eq:cls_probability}
	\rho_c^j = \frac{\exp(-d(E^{qry,j}, \mathcal{P}_c))}{\sum_{i}^C \exp(-d(E^{qry,j}, \mathcal{P}_i))}
\end{equation}

Finally, the cross-entropy loss is computed for the query labels using the class probabilities:

\begin{equation}
	\label{eq:proto_loss}
	\mathcal{L}_{PRO} = - \frac{1}{N} \sum_{j}^N \sum_{i}^{C} Y_i^j \log(\rho_i^j)
\end{equation}

Compared to WeaSeL, ProtoSeg is lighter since it only involves one neural network prediction.
However, it is less reliable to handle large domain shifts since the neural network $\Phi$ never actually learns from the support.
During inference, the procedure is similar to Algorithm \ref{alg:protoseg_step}, but without the loss calculation.

\subsubsection{O-WeaSeL and O-ProtoSeg}

There are two possible issues with the original meta-training.
First, the image selection is done randomly with replacement, which causes some images to be used multiple times while others are not used at all.
It also means that performance on one epoch cannot be compared to another.
Second, the more critical issue is that the training always uses the same number of shots, which is the same as the batch size.
In cases where the batch size is large, the model never learns to utilize a small number of shots.

We address these issues by developing the Omni meta-training as in Algorithm \ref{alg:omni_meta_training}.
The key idea is to transform the original source datasets $\mathcal{S}$ into a new form of dataset $\mathcal{F}$ based on the config $F$.
The config contains many parameters, including the range or options of the number of shots.
The transformation ensures all images in $\mathcal{S}^{sup}$ are used and the duplicates are minimized.
The resulting dataset $\mathcal{F}$ contains predefined batches of support and query sets with a diverse number of shots for the support batches.
In every epoch, all batches in $\mathcal{F}$ are used in a random order.
This ensures that every epoch is comparable, all available data are used, and the model learns to utilize various numbers of shots.

\begin{algorithm}
	\caption{Omni Meta-Training}
	\label{alg:omni_meta_training}
	\begin{algorithmic}
		\Require $\mathcal{S}$, $n_{ep}$, $F$
		\State Initialize $\theta$
		\State Transform $\mathcal{S}$ into $\mathcal{F}$ based on $F$
		\For{$e = 1$ to $n_{ep}$}
		\For{$\mathcal{X}^{sup}$, $\acute{\mathcal{Y}}^{sup}$, $\mathcal{X}^{qry}$, $\mathcal{Y}^{qry}$ in $\mathcal{F}$}
		\State Run Step on $\mathcal{X}^{sup}$, $\acute{\mathcal{Y}}^{sup}$, $\mathcal{X}^{qry}$, $\mathcal{Y}^{qry}$ to obtain $\mathcal{L}$
		\State Update $\theta \leftarrow \theta - \text{opt}(\nabla_{\theta} \mathcal{L})$
		\EndFor
		\EndFor
	\end{algorithmic}
\end{algorithm}

One technical issue with the Omni meta-training is that the number of shots can be larger than the maximum batch size.
To handle this, we modify the meta-training step of WeaSeL and ProtoSeg as shown in Algorithm \ref{alg:o_weasel_step} and Algorithm \ref{alg:o_protoseg_step}.
In both algorithms, the batch pair $(\mathcal{X}^{sup},\mathcal{Y}^{sup})$ is split into smaller batch pairs $(X^{sup},Y^{sup})$ of size $B$.
For WeaSeL, the loss calculation and parameters update are done multiple times for each small batch pair.
Meanwhile, for ProtoSeg, the class prototypes are accumulated via concatenation from all small batch pairs.
Then, the tuned neural network in WeaSeL or accumulated prototypes in ProtoSeg can be used just like those in the original algorithms.

\begin{algorithm}
	\caption{O-WeaSeL Meta-Training Step}
	\label{alg:o_weasel_step}
	\begin{algorithmic}
		\Require $\mathcal{X}^{sup}$, $\acute{\mathcal{Y}}^{sup}$, $\mathcal{X}^{qry}$, $\mathcal{Y}^{qry}$, $\acute{\eta}$
		\For{$(X^{sup},Y^{sup})$ in $(\mathcal{X}^{sup},\mathcal{Y}^{sup})$}
		\State Predict $\hat{Y}^{sup} = \Phi_{\theta}(X^{sup})$
		\State Compute $\mathcal{L}_{SCE}$ from $(\acute{Y}^{sup}, \hat{Y}^{sup})$ using \eqref{eq:sce_loss}
		\State Update $\theta \leftarrow \theta - \acute{\eta} \nabla_{\theta} \mathcal{L}_{SCE}$
		\EndFor
		\State Predict $\hat{Y}^{qry} = \Phi_{\theta}(X^{qry})$
		\State Compute $\mathcal{L}_{CE}$ from $(Y^{qry}, \hat{Y}^{qry})$ using \eqref{eq:ce_loss}
		\State \Return $\mathcal{L}_{CE}$
	\end{algorithmic}
\end{algorithm}

\begin{algorithm}
	\caption{O-ProtoSeg Meta-Training Step}
	\label{alg:o_protoseg_step}
	\begin{algorithmic}
		\Require $\mathcal{X}^{sup}$, $\acute{\mathcal{Y}}^{sup}$, $\mathcal{X}^{qry}$, $\mathcal{Y}^{qry}$
		\For{$X^{sup}$ in $\mathcal{X}^{sup}$}
		\State Obtain $E^{sup}$ from $X^{sup}$ using \eqref{eq:embedding}
		\State Accumulate $E^{sup}$ into $\mathcal{E}^{sup}$
		\EndFor
		\State Compute $\mathcal{P}_c$ for each $c$ from $(\mathcal{E}^{sup}, \acute{\mathcal{Y}}^{sup})$ using \eqref{eq:cls_prototype}
		\State Compute $\rho_c$ for each $c$ from $(\mathcal{X}^{qry}, \mathcal{P}_c)$ using \eqref{eq:cls_probability}
		\State Compute $\mathcal{L}_{PRO}$ from $(\mathcal{Y}^{qry}, \rho_c)$ using \eqref{eq:proto_loss}
		\State \Return $\mathcal{L}_{PRO}$
	\end{algorithmic}
\end{algorithm}

\subsubsection{EO-WeaSeL and EO-ProtoSeg}

The O-WeaSeL is inefficient since it involves multiple operations for each small batch pair.
To address this, we develop the EO-WeaSeL, whose meta-training step is shown in Algorithm \ref{alg:eo_weasel_step}.
The idea is to accumulate the loss $\mathcal{L}_{SCE}$ from all small batch pairs into the total loss $\mathcal{L}_{TSCE}$.
Then, the neural network $\Phi$ is updated only once per step using the total loss $\mathcal{L}_{TSCE}$.

\begin{algorithm}
	\caption{EO-WeaSeL Meta-Training Step}
	\label{alg:eo_weasel_step}
	\begin{algorithmic}
		\Require $\mathcal{X}^{sup}$, $\acute{\mathcal{Y}}^{sup}$, $\mathcal{X}^{qry}$, $\mathcal{Y}^{qry}$, $\acute{\eta}$
		\For{$(X^{sup},Y^{sup})$ in $(\mathcal{X}^{sup},\mathcal{Y}^{sup})$}
		\State Predict $\hat{Y}^{sup} = \Phi_{\theta}(X^{sup})$
		\State Compute $\mathcal{L}_{SCE}$ from $(\acute{Y}^{sup}, \hat{Y}^{sup})$ using \eqref{eq:sce_loss}
		\State Accumulate $\mathcal{L}_{SCE}$ into $\mathcal{L}_{TSCE}$
		\EndFor
		\State Update $\theta \leftarrow \theta - \acute{\eta} \nabla_{\theta} \mathcal{L}_{TSCE}$
		\State Predict $\hat{Y}^{qry} = \Phi_{\theta}(X^{qry})$
		\State Compute $\mathcal{L}_{CE}$ from $(Y^{qry}, \hat{Y}^{qry})$ using \eqref{eq:ce_loss}
		\State \Return $\mathcal{L}_{CE}$
	\end{algorithmic}
\end{algorithm}

Meanwhile, ProtoSeg and O-ProtoSeg also contain inefficiency that also constrains them.
Note that Equation \eqref{eq:cls_prototype} returns $C$ prototypes, each with shape $B \times M$.
It means that the query embeddings and images should have the same batch size $B$, otherwise an error will occur.

This is a significant constraint that can be addressed by using a least common multiple (LCM) approach.
The prototypes and embedding vectors are repeated so their batch sizes are the LCM of their original batch sizes.
The distance function can then be applied to them.
However, the LCM approach is inefficient since it involves much memory and computation.

A simple alternative is to create prototypes with shape $1 \times M$.
It can be done by simply averaging the prototypes across the batch dimension:

\begin{equation}
	\label{eq:avg_cls_prototype}
	\mathcal{P}_c = \frac{1}{BN_c} \sum_{k}^B \sum_{j}^N Y_c^{jk} \odot E^{jk}
\end{equation}

With this Equation \eqref{eq:avg_cls_prototype} replacing Equation \eqref{eq:cls_prototype}, EO-ProtoSeg can be more efficient and less constrained compared to O-ProtoSeg.

\section{Experiments}

\subsection{Datasets}

There are three datasets used in this work: REFUGE \cite{fuREFUGERetinalFundus2019,orlandoREFUGEChallengeUnified2020}, DRISHTI-GS \cite{sivaswamyDrishtiGSRetinalImage2014}, and RIM-ONE r3 \cite{fumeroInteractiveToolDatabase2015}.
They all contain dense labels from multiple experts for OD and OC segmentation.
These datasets are chosen because they are widely used in previous works \cite{liuCADAMultiscaleCollaborative2022,zhangConvolutionalAutoencoderJoint2022,zhouUnsupervisedDomainAdaptation2024,chenIOSUDAUnsupervisedDomain2021,heSelfensemblingMaskboundaryDomain2024,wangPatchBasedOutputSpace2019,bianUnsupervisedOpticDisc2019,liuECSDNetJointOptic2022,xuMinimizingEntropyFourierConsistency2021,kadambiWGANDomainAdaptation2020,zhangUnsupervisedDomainAdaptation2022,wuMinimizingEstimatedRisks2022,tangConsistencyAdversarialSemisupervised2023,luWeaklySupervisedSemiSupervised2020} and have different characteristics.

REFUGE \cite{fuREFUGERetinalFundus2019,orlandoREFUGEChallengeUnified2020} or "Retinal Fundus Glaucoma Challenge" is part of Ophthalmic Medical Image Analysis (OMIA) on the Medical Image Computing and Computer Assisted Intervention (MICCAI) 2018.
It was retrieved from various sources, including hospitals and clinical studies.
Its labeling is performed by seven glaucoma experts from the Zhongshan Ophthalmic Center.
Their labels are combined by majority voting and reviewed by a senior glaucoma expert.

DRISHTI-GS \cite{sivaswamyDrishtiGSRetinalImage2014} is obtained from patients of the Aravind Eye Hospital.
Its labeling is performed by four glaucoma experts.
We then combine their labels ourselves by majority voting.

RIM-ONE r3 \cite{fumeroInteractiveToolDatabase2015} is the third version of the Retinal IMage Database for Optic Nerve Evaluation (RIM-ONE).
Its labeling is performed by two experts using a custom tool named DCSeg.
The dataset also contains combined labels from averaging the labels of the two experts.

\begin{figure}[htbp!]
	\centering
	\includegraphics[scale=0.25]{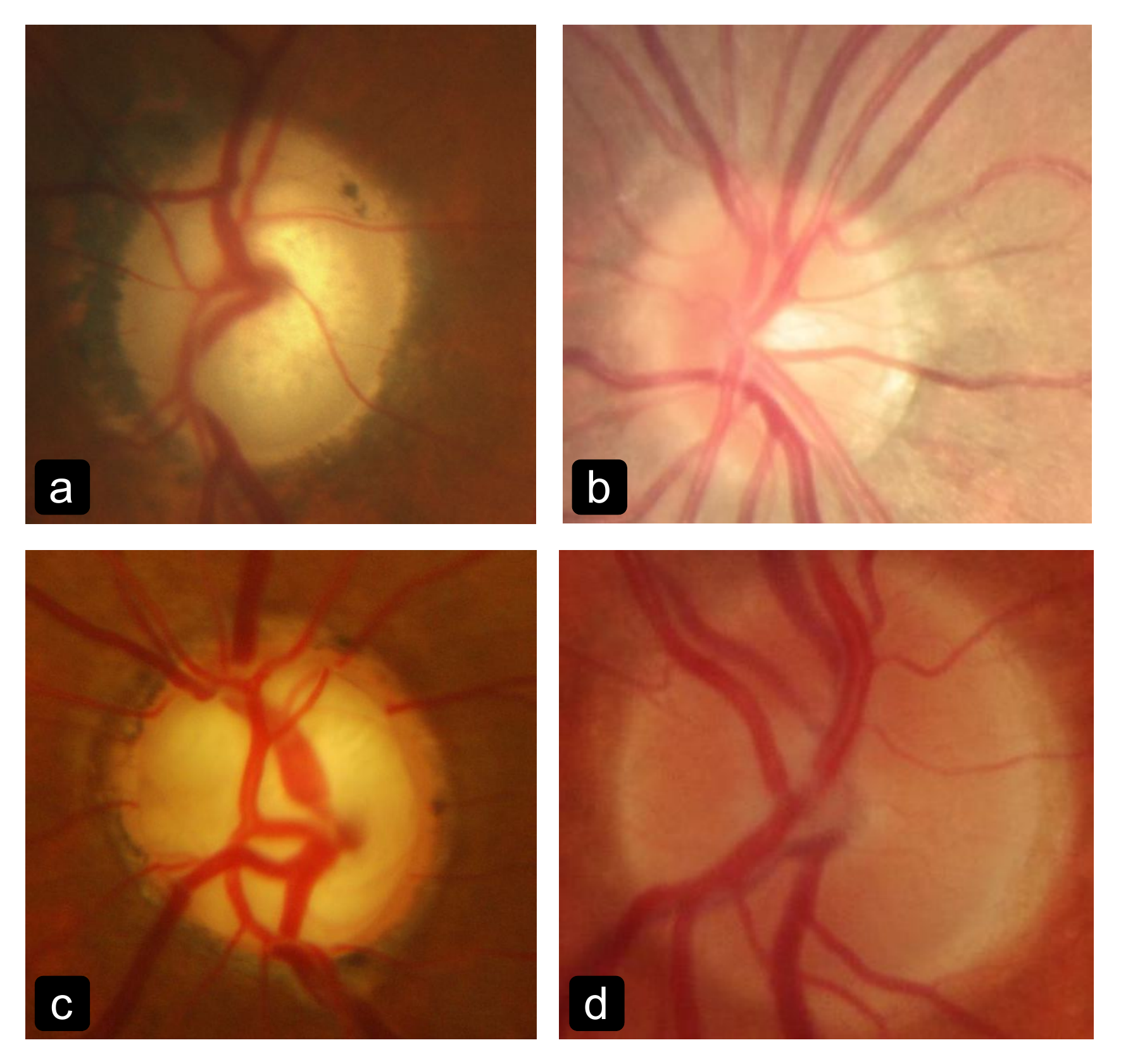}
	\caption{Fundus image sample from each dataset: a. REFUGE train, b. REFUGE val+test, c. DRISHTI-GS, d.~RIM-ONE~r3}
	\label{fig:datasets}
\end{figure}

As shown in Table \ref{tab:dataset}, the datasets have different acquisition techniques regarding camera, field of view (FOV), and resolution.
The differences can also be seen in Fig. \ref{fig:datasets}, which shows a fundus image sample from each dataset.
For the REFUGE dataset, the train images are acquired using Zeiss Visucam 500 while the validation (val) and test images are acquired using Canon CR-2.
Due to this difference, it is common to assume that the REFUGE train and REFUGE val+test are like different datasets.

\begin{table}[htbp!]
	\renewcommand{\arraystretch}{1.2}
	\caption{Dataset acquisition details, class distribution, and mean$\pm$std of padding for cropping}
	\label{tab:dataset}
	\centering
	\begin{tabular}{l l lll l ll l ll}
		\hline
		\multirow{2}{*}{Dataset}                                                           &  & \multicolumn{3}{l}{Acquisition} &            & \multicolumn{2}{l}{Class} &  & \multicolumn{2}{l}{Crop (mean$\pm$std)}                                          \\
		\cline{3-5} \cline{7-8} \cline{10-11}
		                                                                                   &  & Camera                          & FOV        & Resolution                &  & G                                       & N   &  & V Pad         & H Pad         \\
		\hline
		REFUGE train \cite{fuREFUGERetinalFundus2019,orlandoREFUGEChallengeUnified2020}    &  & Zeiss Visucam 500               & -          & 2124$\times$2056          &  & 40                                      & 360 &  & 0.20$\pm$0.08 & 0.27$\pm$0.11 \\
		REFUGE val+test \cite{fuREFUGERetinalFundus2019,orlandoREFUGEChallengeUnified2020} &  & Canon CR-2                      & -          & 1634$\times$1634          &  & 80                                      & 720 &  & 0.26$\pm$0.11 & 0.35$\pm$0.15 \\
		DRISHTI-GS \cite{sivaswamyDrishtiGSRetinalImage2014}                               &  & -                               & 30$^\circ$ & 2896$\times$1944          &  & 70                                      & 31  &  & 0.23$\pm$0.20 & 0.31$\pm$0.27 \\
		RIM-ONE r3 \cite{fumeroInteractiveToolDatabase2015}                                &  & Kowa WX 3D                      & 34$^\circ$ & 2144$\times$1424          &  & 74                                      & 85  &  & 0.17$\pm$0.05 & 0.23$\pm$0.07 \\
		\hline
	\end{tabular}
\end{table}

On each dataset, the fundus images and their dense labels are cropped to the area of OD.
The cropping is performed by adding vertical and horizontal padding to the tight bounding box of the OD.
The padding values are randomly generated from a normal distribution with mean and standard deviation (std) as shown in Table \ref{tab:dataset}.
To increase variability, the mean and std are set differently for each dataset.
The cropped images and masks are saved as files and are ready to be used in the experiments.

\subsection{Training and Evaluation Procedure}

As in standard machine learning, meta-learning experiments in this work involve train, validation, and test sets.
However, in meta-learning experiments, each set has its own support and query.
The characteristics of the support in each set are different based on their roles.
In the train set, it is designed to be diverse but moderate in size, ensuring that the meta-learner can generalize well without too many images per epoch.
Meanwhile, in the test set, it is designed to be structured and large for obtaining evaluation results on multiple cases.
The support in the validation set is just like that in the test set but smaller, allowing for faster evaluation during training.

Table \ref{tab:train_val_test} shows the parameter details of the support.
In the Omni meta-training, these parameters are part of config $F$, which is used to transform the original dataset $\mathcal{S}$ into a new dataset $\mathcal{F}$.
In the original meta-training, the validation and test sets are the same as in Table \ref{tab:train_val_test}, but the train set is different.
As previously mentioned, the original meta-training uses the same number of shots, which is the same as the batch size.
In addition, both the sparse label and its density value are randomized.

\begin{table} [htbp!]
	\renewcommand{\arraystretch}{1.2}
	\caption{Parameter details of the support in the train, validation, and test sets.
	The two numbers inside round brackets "()" represent the low and high limits of a range.
	Meanwhile, the multiple numbers inside square brackets "[]" represent options.}
	\label{tab:train_val_test}
	\centering
	\begin{tabular}{l lll}
		\hline
		         & train      & validation                    & test                         \\
		\hline
		Dataset  & REF train  & REF val, DGS train, RO3 train & REF test, DGS test, RO3 test \\
		Mode     & mix        & combine                       & full combine                 \\
		Shots    & (1, 20)    & [5, 10, 15]                   & [1, 5, 10, 15, 20]           \\
		Points   & (5, 50)    & [13, 25, 37]                  & [1, 13, 25, 37, 50]          \\
		Grid     & (0.1, 1.0) & [0.25, 0.5, 0.75]             & [0.1, 0.25, 0.5, 0.75, 1.0]  \\
		Contours & (0.1, 1.0) & [0.25, 0.5, 0.75]             & [0.1, 0.25, 0.5, 0.75, 1.0]  \\
		Skeleton & (0.1, 1.0) & [0.25, 0.5, 0.75]             & [0.1, 0.25, 0.5, 0.75, 1.0]  \\
		Regions  & (0.1, 1.0) & [0.25, 0.5, 0.75]             & [0.1, 0.25, 0.5, 0.75, 1.0]  \\
		\hline
	\end{tabular}
\end{table}

The datasets used are REFUGE (REF), DRISHTI-GS (DGS), and RIM-ONE r3 (RO3), with their train, validation, and test sets predefined and commonly used.
The "mix" mode in the table means that the numbers of shots and density values are randomly selected from the ranges.
However, it is ensured that the random selection is reproducible and the selection of sparsification is balanced.
The "combine" mode means that each of the possible combinations of numbers of shots, sparse label types, and density values is evaluated once.
Meanwhile, the "full combine" mode means that each of these combinations is evaluated on all query images.
The last five rows are the five sparse labels along with their density value ranges and options.

In addition to the meta-learning experiments, we perform standard fully supervised learning (SL) experiments for comparison.
The first experiment utilizes REF val as the train \& validation sets with REF test as the test set.
In the second one, DGS train and DGS test are used as train \& validation and test sets, respectively.
Similarly, the RO3 train and RO3 test are used in the third one.
A mini-UNet with less than 2 million parameters is used in the SL.
Note that the mini-UNet is also used as the neural network  $\Phi$ for the meta-learners.

All models in meta-learning and SL experiments are trained using the Adam optimizer \cite{KingBa15} with step learning rate scheduler and cross-entropy loss.
Meanwhile, the evaluation metric is Intersection over Union (IoU), which is commonly used and has the lowest bias among other image segmentation metrics \cite{tahaFormalMethodSelecting2014}.
As in previous works \cite{chenReconstructionDrivenDynamicRefinement2023,leiUnsupervisedDomainAdaptation2022,liFewShotDomainAdaptation2021,zhaoWeaklySupervisedSimultaneousEvidence2019,zhaoApplicationAttentionUNet2021}, the metric is calculated for OD and OC separately.
Given binary prediction $\hat{Y}$ and ground truth $Y$, the IoU calculation:

\begin{equation}
	\label{eq:iou}
	IoU = \frac{Y \cap \hat{Y}}{Y \cup \hat{Y}}
\end{equation}

Each method has its hyperparameters optimized using the validation IoU score as the objective.
The values optimized include learning rate, optimizer weight decay and betas, scheduler gamma, inner learning rate and tune epochs for WeaSeL learners, and embedding depth for ProtoSeg learners.

After the optimization, the best model weights are used to evaluate the test set.
While the validation is done on three datasets combined, the test is done on each dataset separately.

\subsection{Profiling Procedure}

Profiling is done to analyze the inference time of each method.
It should be noted that meta-learners have overhead time when learning from the support of the target dataset.
Thus, the inference time of meta-learners cannot be directly compared to that of the SL.
In addition to the overhead time, the inference time includes the prediction time of some query images and the metric calculation time.

Meta-learners can be compared to each other in terms of inference time.
However, WeaSeL learners and ProtoSeg learners should have separate comparisons since the inference time of the former is much larger.
That is because WeaSeL learners involve multiple epochs of tuning while ProtoSeg learners simply generate prototypes.
For extensive profiling, we measure the inference time on the whole REFUGE dataset multiple times, each with a different number of shots and batch size.

While the inference time of meta-learners cannot be directly compared to that of the SL, we can compare the prediction time per image, excluding the overhead time and metric calculation time.
For this, we measure the prediction time of each method on the whole REFUGE dataset multiple times with various batch sizes.

\subsection{Implementation Details}

We use PyTorch \cite{paszkePyTorchImperativeStyle2019} and PyTorch Lightning \cite{falconPyTorchLightningPytorchlightning0762020} for the implementation of meta-learning and SL models and experiments.
Meanwhile, the sparsification techniques are implemented using numpy \cite{harrisArrayProgrammingNumPy2020} and scikit-image \cite{vanderwaltScikitimageImageProcessing2014}.
Our code is available on \href{https://github.com/pandegaabyan/few-shot-weakly-seg}{github.com/pandegaabyan/few-shot-weakly-seg}.
We write the code in an object-oriented manner, as in PyTorch Lightning, with various modules, including data, model, learner, runner, and task.

We have a custom PyTorch dataset class for processing the data during experiments.
It can perform data loading, data splitting \& shuffling, resizing to a fixed size, and conversion to tensors.
Related to FWS, it can also perform sparsification of the dense labels.
The transformation from dataset $\mathcal{S}$ to dataset $\mathcal{F}$ is also performed by the dataset class.

We use Optuna \cite{akibaOptunaNextgenerationHyperparameter2019} for the hyperparameter optimization with Tree-structured Parzen Estimator (TPE) sampler \cite{bergstraAlgorithmsHyperParameterOptimization2011} and Hyperband pruner \cite{liHyperbandNovelBanditbased2017}.
According to Ozaki et al. \cite{ozakiHyperparameterOptimizationMethods2020}, TPE is the best sampler for deep learning cases on limited compute resources and high-dimensional search spaces.
Meanwhile, Hyperband is the best pruner for the TPE sampler, according to the Optuna documentation.
For tracking and logging during the optimization and evaluation, we use Weights and Biases (WandB) \cite{biewaldExperimentTrackingWeights2020}.

Regarding profiling, we modified an existing PyTorch Lightning profiler to make it more suitable but still compatible with the PyTorch Lightning workflow.
The profiler can profile any part of the code using a simple context manager.
The profiling is done on a machine equipped with an NVIDIA Tesla V100 GPU with 32 GB of memory and an Intel Xeon E5 v4 CPU.

\section{Results and Discussion}

In this work, for each meta-learner, there are many experiments with various numbers of shots, sparse label types, and density values on multiple images from three datasets.
The SL experiments are also done on multiple images from the same datasets.
From these extensive experiments, we can analyze the performance of each method in various datasets.
We can also analyze the performance of each method in various numbers of shots and sparse label types.
Lastly, we can compare the results with those from previous works.

\subsection{Analysis of Meta-Learners}

\begin{figure}[htbp!]
	\centering
	\begin{subfigure}[b]{0.43\textwidth}
		\centering
		\includegraphics[scale=0.43]{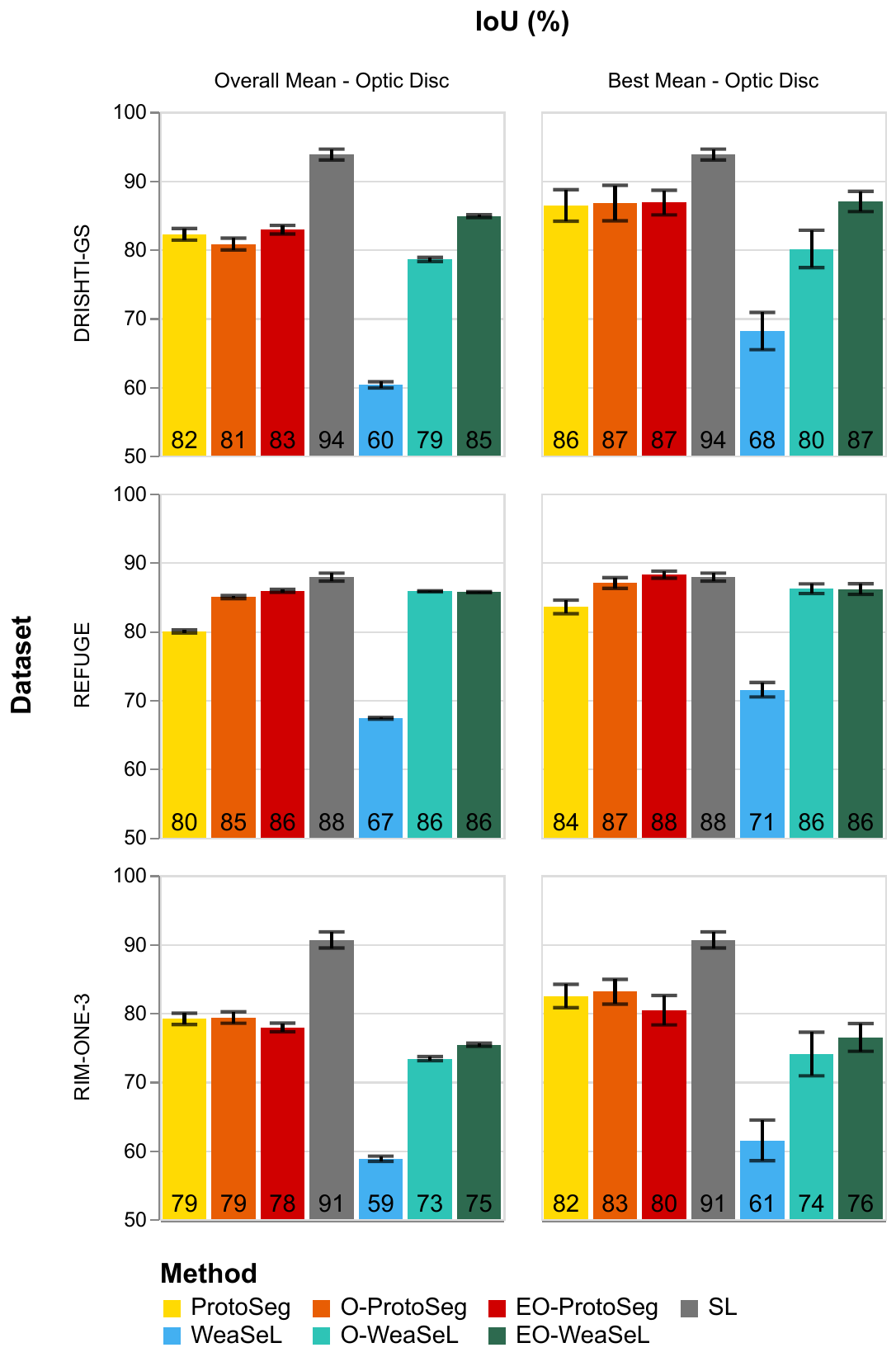}
		\caption{}
		\label{fig:methods_disc}
	\end{subfigure}
	\hspace{1cm}
	\begin{subfigure}[b]{0.43\textwidth}
		\centering
		\includegraphics[scale=0.43]{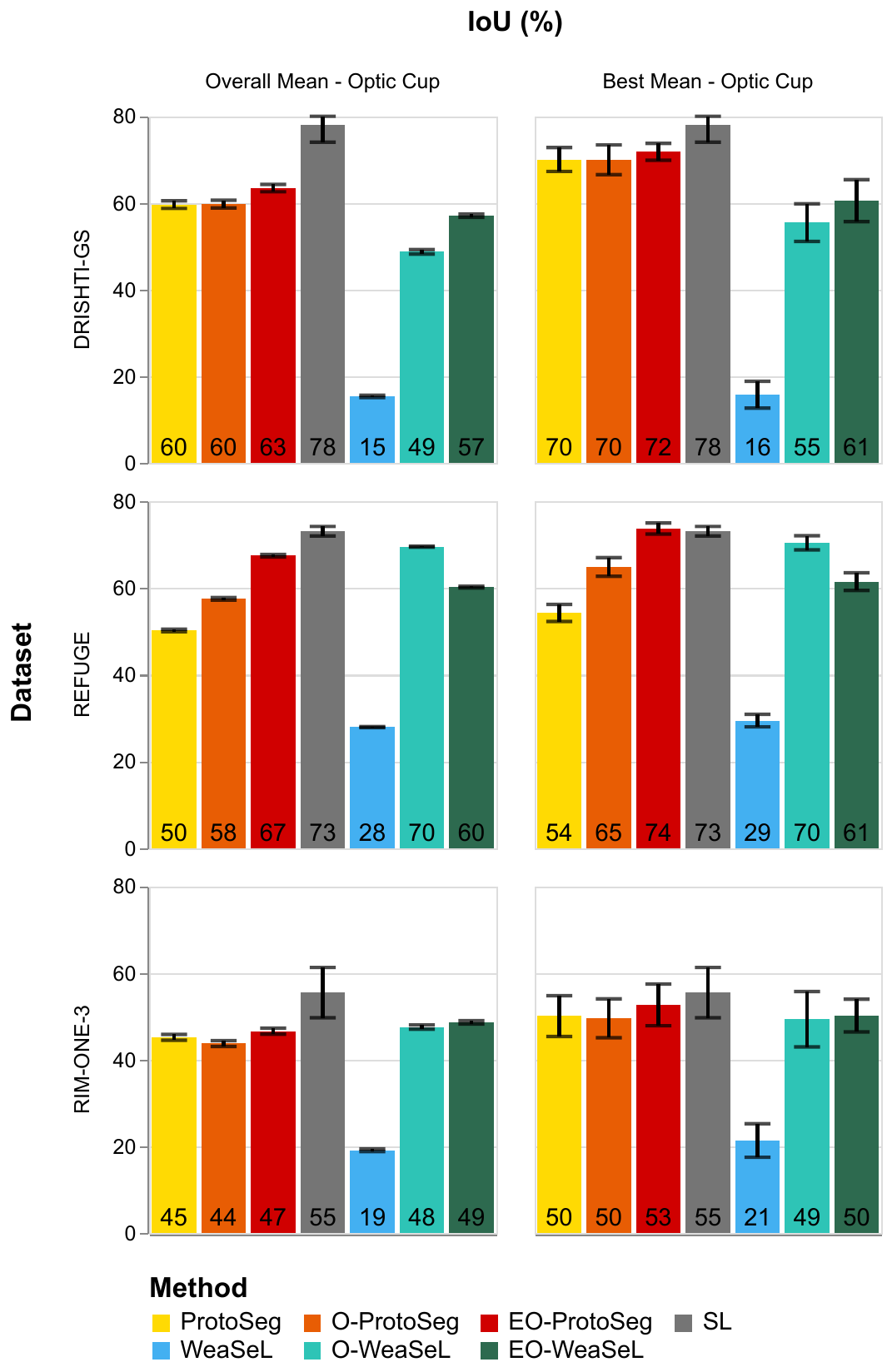}
		\caption{}
		\label{fig:methods_cup}
	\end{subfigure}
	\caption{IoU scores of OD (a) and OC (b) segmentation on various methods and datasets.
		The overall means are averaged across all numbers of shots, sparse label types, and density values. Meanwhile, the best means use a specific number of shots, sparse label type, and density value giving the highest mean IoU score.
		The error bars represent the 95\% confidence interval.}
	\label{fig:methods}
\end{figure}

Fig. \ref{fig:methods} shows the IoU scores of OD and OC segmentation on various methods and datasets.
The overall mean of a method on a dataset is obtained by averaging the IoU scores across all numbers of shots, sparse label types, and density values.
Meanwhile, the best mean is obtained by averaging the IoU scores on a specific number of shots, sparse label type, and density value giving the highest mean IoU score.
A mean IoU score is the simple average of OD IoU score and OC IoU score.
This best combination is different for each method and dataset.
It should be noted that the overall means and the best means of the SL are the same since it does not involve the number of shots or sparse labels.

From Fig. \ref{fig:methods}, it can be seen that the scores of OD and OC share similar trends, in which the shapes of the charts are similar.
The scores of OC are much lower than those of OD, but this is expected since the OC is smaller and more challenging to segment.
The SL results, which become the standard, also show lower OC scores than OD scores.
The confidence intervals of the OC scores are generally wider than those of the OD scores.
Since the number of data points is the same, this means that the OC scores have higher variance, further indicating that OC segmentation is more challenging.

It can also be seen that the best means have similar trends to the overall means.
Obviously, the best means are higher, but the average differences are only around 3\%.
The confidence intervals of the best means are wider, but this is because there are fewer data points.

Regarding the datasets, most meta-learners achieve the best results on the REFUGE dataset.
This is likely because the training set is REF train, which may share some characteristics with REF test, although they look different visually.
Another possible reason is that the REF val dominates the validation set, so the model with the best validation score may perform best on the REF test.
However, the REFUGE dataset still has the best results even after balancing the validation set.
While it is unclear whether REFUGE is the "easiest" dataset, RIM-ONE r3 is clearly the "hardest" dataset.
All methods, including the SL, perform worse on RIM-ONE r3 than the other datasets.

As for the meta-learners, O-WeaSeL and EO-WeaSeL always perform significantly better than the original WeaSeL.
On the REFUGE dataset, O-ProtoSeg and EO-ProtoSeg also perform significantly better than the original ProtoSeg.
Meanwhile, the three ProtoSeg learners have no significant difference on other datasets.
In almost all cases, the Omni learners and their efficient versions have no significant difference.
The exception is WeaSeL on the DRISHTI-GS dataset, where EO-WeaSeL performs significantly better than O-WeaSeL.
It should be noted that EO-ProtoSeg performs worse on the RIM-ONE r3 dataset compared to other ProtoSeg learners, but the difference is not very significant.
With these results, we can conclude that the Omni versions of both WeaSeL and ProtoSeg are better than the original versions.
The balanced data usage and various numbers of shots are proven to be beneficial, especially for WeaSeL.
We can also conclude that the more efficient computation of EO-WeaSeL and EO-ProtoSeg does not hurt the performance.
EO-WeaSeL even performs better than O-WeaSeL in some cases.

In general, WeaSeL versions perform worse than ProtoSeg versions.
While WeaSeL is better at handling large domain shifts \cite{gamaWeaklySupervisedFewShot2023}, we notice that it often fails to learn appropriately from the support during tuning.
Despite the use of hyperparameter optimization, finding the correct inner learning rate and number of epochs for tuning is still challenging.
A low inner learning rate and a high number of epochs can lead to overfitting, while a high inner learning rate and a low number of epochs can lead to underfitting.
More importantly, there is no way to tell whether the tuning overfits or underfits, as there is no validation set during tuning.

Meanwhile, ProtoSeg is less reliable in handling large domain shifts \cite{gamaWeaklySupervisedFewShot2023}, but these experiments do not involve large domain shifts.
Prototype-based methods like ProtoSeg are known to be reliable for few-shot learning \cite{changFewshotSemanticSegmentation2023}.
Other novel methods used in FWS also involve prototypes \cite{langBaseMetaNew2023,chengHolisticPrototypeActivation2023,hanPseudolabelingBasedWeakly2024}.
Given WeaSeL's problems, the nature of these experiments, and the reliability of prototypes, it is not surprising that ProtoSeg generally performs better than WeaSeL.

When comparing meta-learners with SL, the overall means are less suitable since they consist of results from various cases.
The best means are preferred since they are specific and represent the best way to use the methods.
For a more detailed comparison, Table \ref{tab:best_methods} shows the best means of each method on each dataset along with the number of shots, sparse label type, density value, and confidence intervals.
While some meta-learners can achieve results close to those of the SL, generally the SL still performs significantly better.
The differences between best meta-learner results and SL results are around 2\%, 5\%, and 7\% for REFUGE, DRISHTI-GS, and RIM-ONE r3 datasets, respectively.

\begin{table}[htbp!]
	\renewcommand{\arraystretch}{1.2}
	\caption{Best means of each method on each dataset with their parameters and confidence intervals.
		Note that the best means use a specific number of shots, sparse label type, and density value giving the highest mean IoU score.
		The bold rows represent the best method in each dataset.}
	\label{tab:best_methods}
	\centering
	\begin{tabular}{ll l ll l ll l ll}
		\hline
		\multirow{2}{*}{Dataset}    & \multirow{2}{*}{Method} &  & \multicolumn{2}{l}{Parameters} &                          & \multicolumn{2}{l}{OD} &                & \multicolumn{2}{l}{OC}                                            \\
		\cline{4-5} \cline{7-8} \cline{10-11}
		                            &                         &  & Shots                          & Sparse Label             &                        & Mean           & 95\% CI                &  & Mean           & 95\% CI              \\
		\hline
		\multirow{7}{*}{DRISHTI-GS} & ProtoSeg                &  & 20                             & regions (1.00)           &                        & 86.39          & 84.10-88.68            &  & 70.02          & 67.28-72.77          \\
		                            & O-ProtoSeg              &  & 20                             & regions (1.00)           &                        & 86.73          & 84.15-89.31            &  & 69.94          & 66.52-73.36          \\
		                            & \textbf{EO-ProtoSeg}    &  & \textbf{15}                    & \textbf{points (50)}     &                        & \textbf{86.80} & \textbf{85.01-88.59}   &  & \textbf{71.78} & \textbf{69.83-73.73} \\
		                            & SL                      &  & -                              & -                        &                        & 92.92          & 92.28-93.55            &  & 77.12          & 74.64-79.60          \\
		                            & WeaSeL                  &  & 20                             & regions (0.50)           &                        & 68.12          & 65.42-70.83            &  & 15.77          & 12.68-18.87          \\
		                            & O-WeaSeL                &  & 1                              & regions (0.50)           &                        & 80.06          & 77.33-82.78            &  & 55.47          & 51.12-59.82          \\
		                            & EO-WeaSeL               &  & 5                              & grid (0.50)              &                        & 86.97          & 85.49-88.44            &  & 60.52          & 55.67-65.37          \\
		\hline
		\multirow{7}{*}{REFUGE}     & ProtoSeg                &  & 20                             & contours (1.00)          &                        & 83.52          & 82.54-84.49            &  & 54.25          & 52.27-56.22          \\
		                            & O-ProtoSeg              &  & 20                             & regions (0.50)           &                        & 87.00          & 86.22-87.78            &  & 64.85          & 62.68-67.02          \\
		                            & \textbf{EO-ProtoSeg}    &  & \textbf{20}                    & \textbf{regions (0.75)}  &                        & \textbf{88.21} & \textbf{87.69-88.73}   &  & \textbf{73.70} & \textbf{72.43-74.97} \\
		                            & SL                      &  & -                              & -                        &                        & 90.18          & 89.91-90.46            &  & 75.96          & 75.37-76.56          \\
		                            & WeaSeL                  &  & 20                             & grid (0.25)              &                        & 71.48          & 70.42-72.53            &  & 29.41          & 27.95-30.87          \\
		                            & O-WeaSeL                &  & 20                             & grid (0.25)              &                        & 86.16          & 85.46-86.87            &  & 70.41          & 68.77-72.04          \\
		                            & EO-WeaSeL               &  & 20                             & grid (0.25)              &                        & 86.11          & 85.34-86.88            &  & 61.47          & 59.43-63.51          \\
		\hline
		\multirow{7}{*}{RIM-ONE r3} & ProtoSeg                &  & 10                             & regions (0.25)           &                        & 82.45          & 80.76-84.15            &  & 50.06          & 45.36-54.77          \\
		                            & O-ProtoSeg              &  & 15                             & points (13)              &                        & 83.08          & 81.28-84.87            &  & 49.57          & 45.07-54.06          \\
		                            & \textbf{EO-ProtoSeg}    &  & \textbf{20}                    & \textbf{skeleton (0.10)} &                        & \textbf{80.39} & \textbf{78.26-82.52}   &  & \textbf{52.65} & \textbf{47.83-57.46} \\
		                            & SL                      &  & -                              & -                        &                        & 90.93          & 90.16-91.70            &  & 56.12          & 51.91-60.34          \\
		                            & WeaSeL                  &  & 15                             & skeleton (0.50)          &                        & 61.47          & 58.51-64.43            &  & 21.37          & 17.49-25.24          \\
		                            & O-WeaSeL                &  & 10                             & grid (0.50)              &                        & 74.02          & 70.84-77.19            &  & 49.35          & 42.95-55.74          \\
		                            & EO-WeaSeL               &  & 5                              & regions (0.75)           &                        & 76.44          & 74.42-78.46            &  & 50.19          & 46.43-53.96          \\
		\hline
	\end{tabular}
\end{table}

Based on Table \ref{tab:best_methods}, it is quite clear that the best meta-learner is EO-ProtoSeg since its mean IoU outperforms others on all datasets.
It further validates that the Omni meta-training improves the performance while its efficient version does not hurt the performance.
It can also be seen that the most common number of shots is 20 while the most common sparse label types are regions and grid with varying density values.

\subsection{Analysis of Number of Shots and Sparse Label}

\begin{figure}[htbp!]
	\centering
	\includegraphics[scale=0.5]{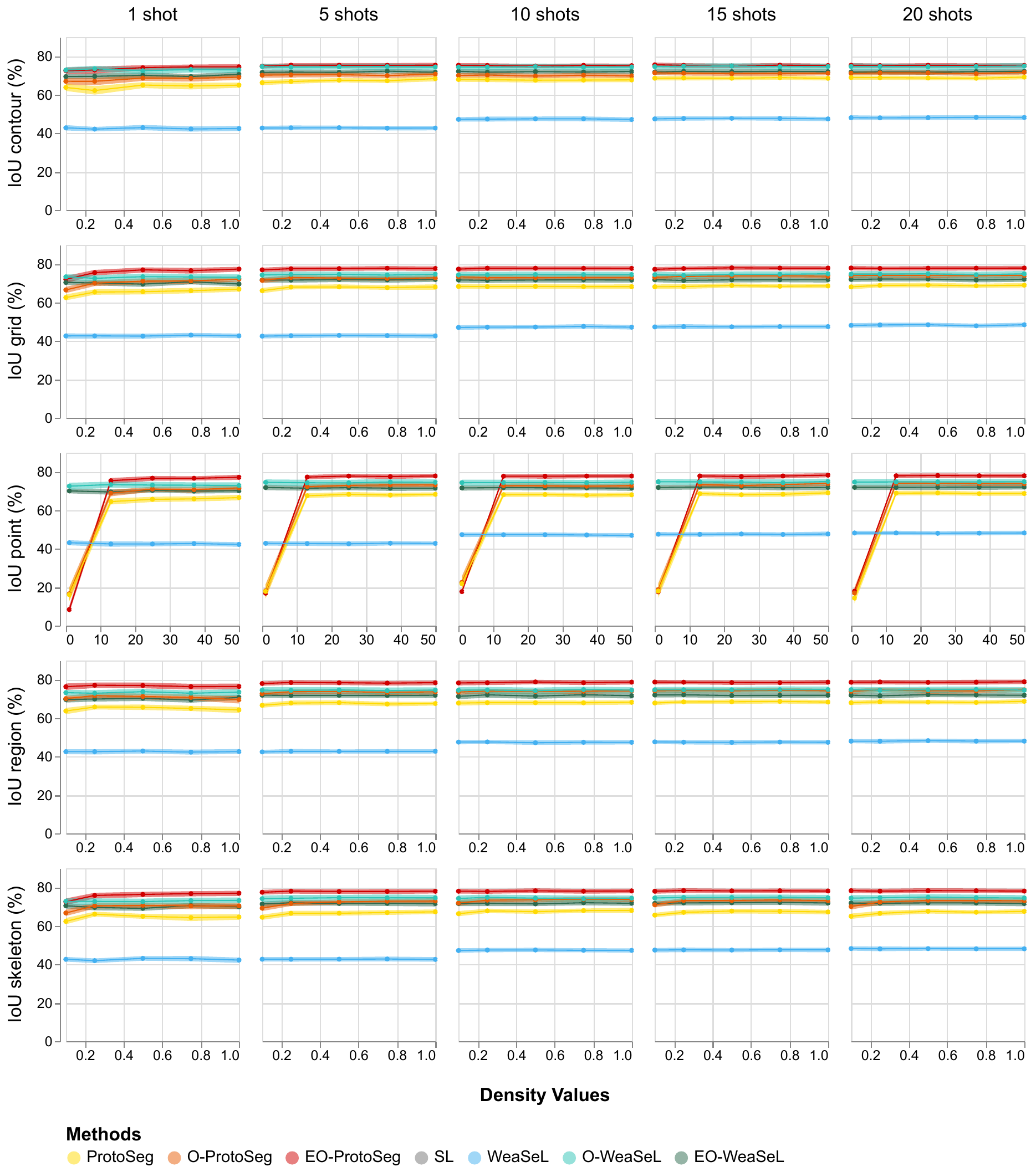}
	\caption{Mean IoU scores of multiple methods on combined datasets with various numbers of shots (top) and sparse labels (bottom).
		The error area represents the 95\% confidence interval.}
	\label{fig:shots_sparse_labels}
\end{figure}

Fig. \ref{fig:shots_sparse_labels} shows the scores of multiple methods with various numbers of shots, sparse label types, and density values.
The scores are the means of OD IoU scores and OC IoU scores across all datasets.
It should be noted that the number of data points in each dataset is not the same, so each dataset contributes differently to the scores.
While small in size, there are error areas that represent the 95\% confidence interval.

In general, the trend looks similar across various sparse label types.
EO-ProtoSeg performs significantly better than other meta-learners in almost all cases, except in contours labels where EO-ProtoSeg is only slightly better than O-WeaSeL.
Meanwhile, in almost all cases, the original WeaSeL performs much worse than other meta-learners.
One important exception is when the labels are just one point, in which WeaSeL learners perform better than EO-ProtoSeg and other ProtoSeg learners.
The lousy performance is understandable since one point means that only one class is labeled with one point while other classes are not labeled at all.

Interestingly, WeaSeL learners can still perform well with one point labels.
WeaSeL learners may utilize more knowledge from the meta-training on the source dataset instead of learning from the support.
This also explains why WeaSeL learners are generally unaffected by density values, as seen from the plateau shapes in the charts.
However, the number of shots still affects the performance of WeaSeL learners, with more shots leading to better performance.
In some cases on original WeaSeL, the performance increase is significant but less significant in most cases.

Regarding ProtoSeg learners, what happens seems to be the opposite.
Density values affect the performance of ProtoSeg learners, with larger density values usually leading to better performance.
Meanwhile, the effect of the number of shots is less apparent, except between one shot and other numbers of shots.
In many one shot cases, the performance difference between the lowest density value and the second lowest density value is quite clear.
This trend is not seen in other numbers of shots.
This is because in a one shot case with a low density value, the available information is minimal.
Thus, any additional information can significantly improve the performance.
In a case with a larger number of shots and/or higher density value, the available information is already sufficient.
Thus, any additional information does not significantly improve the performance.

Based on this analysis, we can conclude that WeaSeL learners are not learning optimally from the support.
Meanwhile, ProtoSeg learners can learn properly but have limited knowledge capacity, limiting their performance.
However, it does not mean that all of them perform poorly.
EO-ProtoSeg can achieve mean IoU scores near 80\% with only one sparsely labeled image.
In comparison, the SL utilizing hundreds of densely labeled images achieves a mean IoU score of 83\% on the REFUGE dataset.

For more detailed results of EO-ProtoSeg as the best meta-learner, Table \ref{tab:eo_protoseg} presents IoU scores of EO-ProtoSeg across various numbers of shots, sparse label types, and density values.
While the difference between sparse label types is unclear in Fig. \ref{fig:shots_sparse_labels}, it is more apparent in Table \ref{tab:eo_protoseg}.
The best sparse label is regions, most likely because it has more labeled pixels.
Despite that, annotating regions is not challenging since annotators only need to pick from predefined regions.
The effectiveness of regions was mentioned by Gama et al. \cite{gamaWeaklySupervisedFewShot2023}, but it depends on the superpixel algorithm used.
Skeleton also performs well due to the lines that can be thick because they are drawn in the center of objects.
Points and grid have similar performance due to their similar characteristics, except in the case of one-point labels, which is explained earlier.
Lastly, contours surprisingly perform worse despite sharing characteristics similar to those of the skeleton.
Since contours are lines drawn on the edges of objects, they are usually thin and thus have fewer labeled pixels than skeleton lines.
Drawing lines on the edges is also more challenging for annotators, making contours the least effective sparse label.

\begin{table}[htbp!]
	\renewcommand{\arraystretch}{1.2}
	\caption{Mean IoU scores of EO-ProtoSeg on combined datasets with various numbers of shots, sparse label types, and density values.
		The bold numbers represent the best score on each sparse label.}
	\label{tab:eo_protoseg}
	\centering
	\begin{tabular}{ll l lllll}
		\hline
		\multicolumn{2}{l}{Sparse Label} &         & \multicolumn{5}{l}{Shots}                                                           \\
		\cline{1-2} \cline{4-8}
		Type                             & Density &                           & 1     & 5     & 10    & 15             & 20             \\
		\hline
		\multirow{5}{*}{contours}        & 0.10    &                           & 72.47 & 74.82 & 75.29 & \textbf{75.53} & 75.28          \\
		                                 & 0.25    &                           & 72.83 & 75.34 & 75.17 & 75.22          & 75.12          \\
		                                 & 0.50    &                           & 74.02 & 75.27 & 74.88 & 74.92          & 75.27          \\
		                                 & 0.75    &                           & 74.46 & 75.25 & 75.22 & 75.33          & 75.12          \\
		                                 & 1.00    &                           & 74.48 & 75.30 & 75.09 & 75.13          & 75.19          \\
		\hline
		\multirow{5}{*}{grid}            & 0.10    &                           & 72.04 & 76.96 & 77.33 & 77.25          & 77.91          \\
		                                 & 0.25    &                           & 75.50 & 77.54 & 77.80 & 77.64          & 77.68          \\
		                                 & 0.50    &                           & 76.94 & 77.62 & 77.79 & \textbf{78.00} & 77.79          \\
		                                 & 0.75    &                           & 76.54 & 77.75 & 77.74 & 77.86          & 77.82          \\
		                                 & 1.00    &                           & 77.37 & 77.69 & 77.72 & 77.87          & 77.89          \\
		\hline
		\multirow{5}{*}{points}          & 1       &                           & 8.41  & 16.90 & 17.73 & 17.62          & 17.95          \\
		                                 & 13      &                           & 75.37 & 77.27 & 77.71 & 77.83          & 77.93          \\
		                                 & 25      &                           & 76.64 & 77.76 & 77.69 & 77.54          & 78.08          \\
		                                 & 37      &                           & 76.62 & 77.52 & 77.76 & 77.84          & 77.95          \\
		                                 & 50      &                           & 77.13 & 77.83 & 77.82 & \textbf{78.23} & 77.94          \\
		\hline
		\multirow{5}{*}{regions}         & 0.10    &                           & 76.30 & 77.96 & 78.27 & 78.73          & 78.63          \\
		                                 & 0.25    &                           & 77.05 & 78.48 & 78.37 & 78.63          & 78.78          \\
		                                 & 0.50    &                           & 76.94 & 78.36 & 78.78 & 78.49          & 78.55          \\
		                                 & 0.75    &                           & 76.34 & 78.16 & 78.39 & 78.41          & 78.65          \\
		                                 & 1.00    &                           & 76.42 & 78.38 & 78.69 & 78.67          & \textbf{78.89} \\
		\hline
		\multirow{5}{*}{skeleton}        & 0.10    &                           & 72.52 & 77.47 & 78.02 & 78.01          & 78.31          \\
		                                 & 0.25    &                           & 75.82 & 78.04 & 77.90 & \textbf{78.37} & 78.05          \\
		                                 & 0.50    &                           & 76.35 & 77.88 & 78.23 & 78.18          & 78.31          \\
		                                 & 0.75    &                           & 76.74 & 77.89 & 78.02 & 78.25          & 78.27          \\
		                                 & 1.00    &                           & 76.97 & 78.00 & 78.17 & 78.13          & 78.08          \\
		\hline
	\end{tabular}
\end{table}

The difference between various numbers of shots is also more apparent in Table \ref{tab:eo_protoseg}.
While all of the best scores are achieved with a large number of shots, some notable good scores are achieved with a small number of shots and low density values.
The second highest score in Table \ref{tab:eo_protoseg} is 78.78\%, achieved with only 10 shots and 0.50 regions density.
This score is only 0.11\% lower than the best score.

The best score for five shots is 78.48\% which is 0.41\% lower than the best score.
It is also from regions with a 0.25 density value.
In the case of one shot, the best scores are 77.37\% with full grid density and 77.13\% with 50 points density.
However, both full grid and 50 point density are hard to annotate.
The best score for one shot with easy annotation is 77.05\% with 0.25 regions density.
This further validates the effectiveness of regions label.

\subsection{Analysis of Inference Time}

\begin{figure}[htbp!]
	\centering
	\includegraphics[scale=0.5]{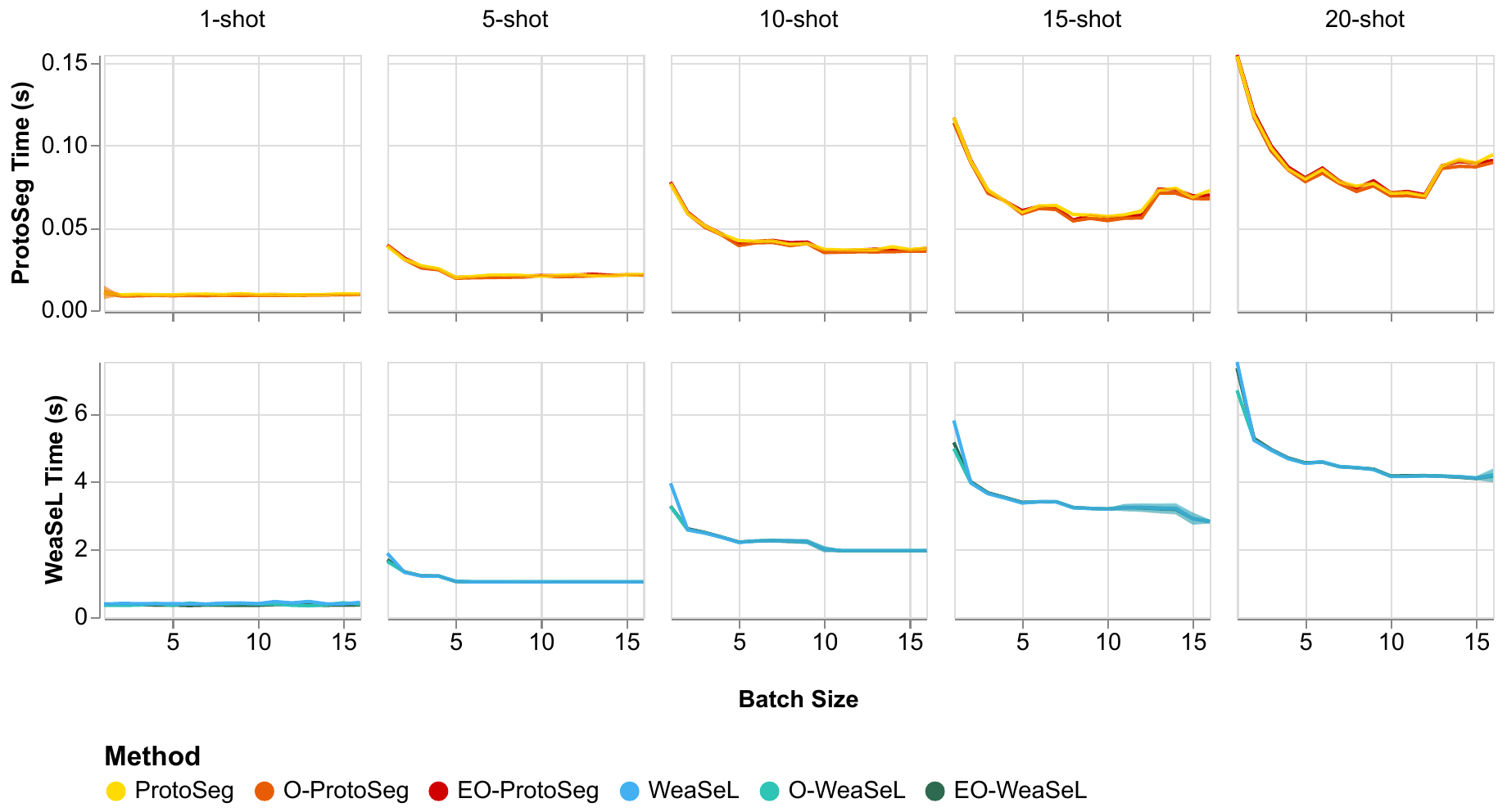}
	\caption{Inference time of ProtoSeg and WeaSeL learners across various numbers of shots and batch sizes.}
	\label{fig:inference_time}
\end{figure}

Figure \ref{fig:inference_time} shows the inference time of WeaSeL and ProtoSeg learners.
There are error areas that represent the 95\% confidence interval.
It can be seen that the larger the number of shots, the longer the inference time.
That is because the overhead time becomes longer when more images are in the support.

In the case of ProtoSeg learners, the original and O-ProtoSeg have similar inference times.
However, in most cases, EO-ProtoSeg has significantly shorter inference times.
This means that averaging the prototypes across the batch dimension gives a significant speedup.
As previously discussed, this modification does not hurt the performance.
Looking at the batch size, larger batch sizes do not always lead to shorter inference time.
It is likely that the overhead process, which is the prototype generation, needs more time when the batch size is too large.

In the case of WeaSeL learners, all versions have similar inference times with no significant difference.
While EO-WeaSeL has fewer updates due to the accumulated loss, its updates may take longer, making the total time similar to the other versions.
The more plausible reason is that the multiple updates of WeaSeL and O-WeaSeL are more efficient than previously thought, likely due to the optimized PyTorch calculation.
Regarding batch size, larger batch sizes mostly lead to shorter inference times.
This is expected since the tuning and prediction in WeaSeL learners resemble standard training and prediction in the SL method, which are known to be faster with larger batch sizes.

\begin{figure}[htbp!]
	\centering
	\includegraphics[scale=0.5]{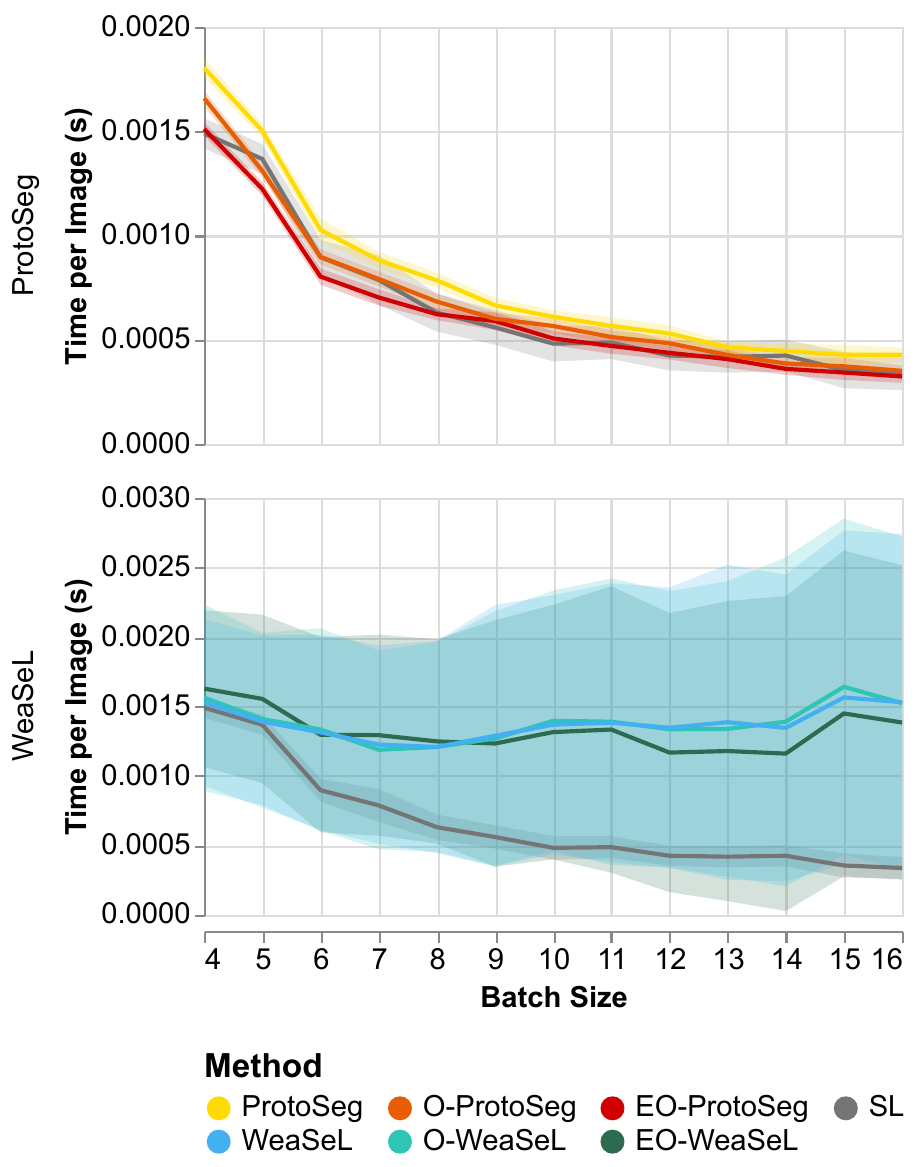}
	\caption{Prediction time per image of ProtoSeg, WeaSeL, and SL methods across various batch sizes.}
	\label{fig:prediction_time}
\end{figure}

Figure \ref{fig:prediction_time} shows the prediction time per image of ProtoSeg, WeaSeL, and SL methods.
While the charts of ProtoSeg and WeaSeL are separated, notice that the SL is included in both charts.

In general, there is no significant difference between methods.
However, EO-ProtoSeg has slightly shorter prediction times compared to other ProtoSeg learners.
EO-ProtoSeg is faster since its averaged prototypes have fewer dimensions.
Meanwhile, the WeaSeL chart shows that WeaSeL learners have high variances in prediction time, with their mean prediction time being higher than that of the SL.

\subsection{Comparison with Previous Works}

Table \ref{tab:comparison} shows the comparison of EO-ProtoSeg with previous works.
Most methods that use the same dataset utilization and perform the same OD and OC segmentation are UDA methods.
While UDA differs from FWS in that it does not require labels in the source dataset, it is the most common approach in the literature.
However, most UDA methods need access to the target dataset during training \cite{zhangRobustColorMedical2022,huDevilChannelsContrastive2023}, which may not be viable.
It also means that each time the target domain is changed, full retraining is needed \cite{wangPatchBasedOutputSpace2019}.

\begin{table}[htbp!]
	\renewcommand{\arraystretch}{1.2}
	\caption{Comparison of the best method in this work with previous works on three datasets.
		The Params column refers to the number of parameters of the backbone neural network of each method.
		The UDA methods have the same dataset utilization as ours where the source dataset is REF train.
		The WSS method utilizes bounding boxes as weak labels.
		The FSS method uses 10 support images while the SSS method uses 160 labeled images.
		EO-ProtoSeg best refers to the best means as in the bold rows of Table \ref{tab:best_methods}.
		Meanwhile, the 1s, 5s, and 10s refer to 1 shot with 0.25 regions density, 5 shots with 0.25 regions density, and 10 shots with 0.50 regions density, respectively.}
	\label{tab:comparison}
	\centering
	\begin{tabular}{l l l l l l ll l ll l ll}
		\hline
		\multirow{2}{*}{Method}                                    &  & \multirow{2}{*}{Approach} &  & \multirow{2}{*}{Params} &  & \multicolumn{2}{l}{DRISHTI-GS} &       & \multicolumn{2}{l}{REFUGE} &       & \multicolumn{2}{l}{RIM-ONE r3}                    \\
		\cline{7-8} \cline{10-11} \cline{13-14}
		                                                           &  &                           &  &                         &  & OD                             & OC    &                            & OD    & OC                             &  & OD    & OC    \\
		\hline
		CFEA \cite{liuCFEACollaborativeFeature2019}                &  & UDA                       &  & -                       &  & 79.78                          & 70.52 &                            & 88.96 & 75.86                          &  & 60.08 & 46.53 \\
		pOSAL \cite{wangPatchBasedOutputSpace2019}                 &  & UDA                       &  & 5.8M                    &  & 91.42                          & 72.30 &                            & 90.83 & 78.31                          &  & 76.75 & 62.59 \\
		SIFA \cite{chenUnsupervisedBidirectionalCrossModality2020} &  & UDA                       &  & 43.3M                   &  & 83.04                          & 57.29 &                            & 85.69 & 69.57                          &  & 74.67 & 52.84 \\
		WGAN \cite{kadambiWGANDomainAdaptation2020}                &  & UDA                       &  & -                       &  & 91.20                          & 72.40 &                            & -     & -                              &  & -     & -     \\
		IOSUDA \cite{chenIOSUDAUnsupervisedDomain2021}             &  & UDA                       &  & 42.8M                   &  & 89.53                          & 65.56 &                            & 91.04 & 71.03                          &  & 83.26 & 60.07 \\
		CADA \cite{liuCADAMultiscaleCollaborative2022}             &  & UDA                       &  & 9.7M                    &  & 80.18                          & 72.41 &                            & 90.44 & 77.21                          &  & 62.13 & 47.1  \\
		SCUDA \cite{zhangUnsupervisedDomainAdaptation2022}         &  & UDA                       &  & -                       &  & 90.34                          & 66.61 &                            & -     & -                              &  & 84.89 & 61.65 \\
		GrabCut + UNet \cite{luWeaklySupervisedSemiSupervised2020} &  & WSS                       &  & -                       &  & 86.37                          & -     &                            & -     & -                              &  & -     & -     \\
		MERU \cite{wuMinimizingEstimatedRisks2022}                 &  & FSS                       &  & -                       &  & -                              & -     &                            & 83.92 & 61.47                          &  & -     & -     \\
		RDMT \cite{meiSemisupervisedImageSegmentation2024}         &  & SSS                       &  & -                       &  & -                              & -     &                            & -     & 70.93                          &  & -     & -     \\
		EO-ProtoSeg 1s (ours)                                      &  & FWS                       &  & 1.9M                    &  & 84.96                          & 63.69 &                            & 88.15 & 71.17                          &  & 79.92 & 44.01 \\
		EO-ProtoSeg 5s (ours)                                      &  & FWS                       &  & 1.9M                    &  & 85.30                          & 68.61 &                            & 88.18 & 73.11                          &  & 80.46 & 50.27 \\
		EO-ProtoSeg 10s (ours)                                     &  & FWS                       &  & 1.9M                    &  & 85.02                          & 68.93 &                            & 88.18 & 73.52                          &  & 80.57 & 52.42 \\
		EO-ProtoSeg best (ours)                                    &  & FWS                       &  & 1.9M                    &  & 86.80                          & 71.78 &                            & 88.21 & 73.70                          &  & 80.39 & 52.65 \\
		\hline
	\end{tabular}
\end{table}

Table \ref{tab:comparison} shows that EO-ProtoSeg has comparable performance with UDA methods and even outperforms non-UDA methods.
While some UDA methods achieve higher scores, they usually have many more parameters than EO-ProtoSeg, with some methods having twenty times more.
EO-ProtoSeg also does not require any retraining when the target domain is changed.
It just needs to extract prototypes from a few labeled images, which can be done in around a tenth of a second.
This makes EO-ProtoSeg more practical and efficient than UDA methods in a limited compute resource environment.

EO-ProtoSeg best outperforms non-UDA methods, although there are not many scores to compare.
Moreover, EO-ProtoSeg 1s, which uses only one image labeled with 0.25 regions density, is better than the FSS method, which uses 10 support images, and the SSS method, which uses 160 labeled images.
This further validates the effectiveness of EO-ProtoSeg in OD and OC segmentation with limited labeled data.

\section{Conclusion}

This study presents meta-learners for few-shot weakly supervised OD and OC segmentation on fundus images.
We improve WeaSeL and ProtoSeg by introducing Omni meta-training, which balances the data usage and diversifies the number of shots.
We also introduce efficient versions of them that reduce the computational cost.

The results show that Omni versions consistently outperform original versions while efficient versions achieve similar performance to Omni versions.
ProtoSeg learners generally perform better than WeaSeL.
The latter often fails to learn optimally from the support while the former can learn with limited knowledge capacity.

Larger shots generally yield better results, but smaller shot scenarios can still perform well with effective sparse labels.
The regions label is the most effective due to its larger but simple annotations.

Profiling indicates that, in most cases, the inference time of EO-ProtoSeg is significantly shorter than that of other ProtoSeg learners.
The prediction time of ProtoSeg learners is comparable to that of the SL.
However, this trend does not hold for WeaSeL learners.

EO-ProtoSeg achieves IoU scores of 88.15\% for OD and 71.17\% for OC on the REFUGE dataset using just one image with 0.25 region density.
It surpasses FSS and SSS methods, which use 10 and 160 labeled images, respectively.
Compared to UDA methods, EO-ProtoSeg is comparable in performance yet more efficient since it is lighter and does not require retraining.

These findings suggest the effectiveness of EO-ProtoSeg and other FWS methods in medical image segmentation with limited labeled data.
With EO-ProtoSeg, a model can be trained optimally and then adapted to new data using only one sparsely labeled image.
Its prediction time is comparable to fully supervised methods while its performance surpasses that of other less-supervised methods, making it a promising solution for real use cases.
Future studies could explore applications to other types of medical images, enhance ProtoSeg's learning capacity, and develop more customizable sparsification techniques to improve annotation efficiency.

\section*{Data and code availability}

All datasets used in this study are publicly available.
The source code used in this study is also publicly available at \url{https://github.com/pandegaabyan/few-shot-weakly-seg}.

\section*{Acknowledgment}

This study was funded by the Ministry of Higher Education, Science, and Technology of Indonesia through the Basic Research schema and the PMDSU program with grant number 048/E5/PG.02.00.PL/2024.

\bibliographystyle{cas-model2-names}

\bibliography{ref}

\end{document}